\documentclass[onecolumn,12pt]{article}
\usepackage{graphicx}
\usepackage{longtable}
\usepackage{lscape}
\usepackage{color}

\newcommand\simlt{\hspace{0.3em}\raisebox{0.4ex}{$<$}\hspace{-0.75em}\raisebox{-.7ex}{$\sim$}\hspace{0.3em}} 
\newcommand\simgt{\hspace{0.3em}\raisebox{0.4ex}{$>$}\hspace{-0.75em}\raisebox{-.7ex}{$\sim$}\hspace{0.3em}} 

\newcommand{\beginsupplement}{%
        \setcounter{table}{0}
        \renewcommand{\thetable}{S\arabic{table}}%
        \setcounter{figure}{0}
        \renewcommand{\thefigure}{S\arabic{figure}}%
     }

\begin{document}
\baselineskip 14pt

\title{Distortion of Magnetic Fields in a Starless Core VI: \\
Application of Flux Freezing Model and \\
Core Formation of FeSt 1-457} 
\date{\underline{Ver.11}}
\author{Ryo Kandori$^{1}$, Kohji Tomisaka$^{2}$, Masao Saito$^{2}$, Motohide Tamura$^{1,2,3}$, \\
Tomoaki Matsumoto$^4$, Ryo Tazaki$^{5}$, Tetsuya Nagata$^{6}$, Nobuhiko Kusakabe$^{1}$, \\
Yasushi Nakajima$^{7}$, Jungmi Kwon$^{3}$, Takahiro Nagayama$^{8}$, and Ken'ichi Tatematsu$^{2}$\\
{\small 1. Astrobiology Center of NINS, 2-21-1, Osawa, Mitaka, Tokyo 181-8588, Japan}\\
{\small 2. National Astronomical Observatory of Japan, 2-21-1 Osawa, Mitaka, Tokyo 181-8588, Japan}\\
{\small 3. Department of Astronomy, The University of Tokyo, 7-3-1, Hongo, Bunkyo-ku, Tokyo, 113-0033, Japan}\\
{\small 4. Faculty of Sustainability Studies, Hosei University, Fujimi, Chiyoda-ku, Tokyo 102-8160}\\
{\small 5. Astronomical Institute, Graduate School of Science Tohoku University,}\\
{\small 6-3 Aramaki, Aoba-ku, Sendai 980-8578, Japan}\\
{\small 6. Kyoto University, Kitashirakawa-Oiwake-cho, Sakyo-ku, Kyoto 606-8502, Japan}\\
{\small 7. Hitotsubashi University, 2-1 Naka, Kunitachi, Tokyo 186-8601, Japan}\\
{\small 8. Kagoshima University, 1-21-35 Korimoto, Kagoshima 890-0065, Japan}\\
{\small e-mail: r.kandori@nao.ac.jp}}
\maketitle

\abstract{Observational data for the hourglass-like magnetic field toward the starless dense core FeSt 1-457 were compared with a flux freezing magnetic field model (Myers et al. 2018). Fitting of the observed plane-of-sky magnetic field using the flux freezing model gave a residual angle dispersion comparable with the results based on a simple three-dimensional parabolic model. The best-fit parameters for the flux freezing model were a line-of-sight magnetic inclination angle of $\gamma_{\rm mag} = 35^{\circ} \pm 15^{\circ}$ and a core center to ambient (background) density contrast of $\rho_{\rm c} / \rho_{\rm bkg} = 75$. The initial density for core formation ($\rho_0$) was estimated to be $\rho_{\rm c} / 75 = 4670$ cm$^{-3}$, which is about one order of magnitude higher than the expected density ($\sim 300$ cm$^{-3}$) for the inter-clump medium of the Pipe Nebula. FeSt 1-457 is likely to have been formed from the accumulation of relatively dense gas, and the relatively dense background column density of $A_V \simeq 5$ mag supports this scenario. The initial radius (core formation radius) $R_0$ and the initial magnetic field strength $B_0$ were obtained to be 0.15 pc ($1.64 R$) and $10.8-14.6$ $\mu$G, respectively. We found that the initial density $\rho_0$ is consistent with the mean density of the nearly critical magnetized filament with magnetic field strength $B_0$ and radius $R_0$. The relatively dense initial condition for core formation can be naturally understood if the origin of the core is the fragmentation of magnetized filaments. 
}
\vspace*{0.3 cm}

\section{Introduction}
The characteristics of newborn stars are thought to be determined by the physical properties of the nursing molecular cloud cores (dense cores). Revealing the formation mechanism of cores is important because it will help determine the initial conditions of star formation. 
\par
Cores are thought to develop and evolve in molecular clouds via a mass accumulation process involving gravity, thermal pressure, turbulence, and magnetic field. Several scenarios have been proposed for the formation mechanism of cores. One is the quasi-static contraction of material under a relatively strong magnetic field (Shu 1977; Shu, Adams, \& Lizano 1987). The other extreme is core formation through supersonic turbulence (e.g., Mac Low \& Klessen 2004). In this scenario, supersonic turbulence produces cores that collapse dynamically, accompanied by highly supersonic infalling motion. However, these models do not match observations in several aspects. Many observations show a moderately supercritical condition in molecular clouds (e.g., Crutcher et al. 2004), which is not the case for the first model. Also, quiescent kinematic gas motions are widely observed toward dense cores (e.g., Caselli et al. 2002), which does not match the second model. A core formation mechanism between those two extreme models may better account for the observations (e.g., Nakamura \& Li 2005; Basu et al. 2009a,b). 
\par
Many observations of dense cores have been made, using various methods at various wavelengths, e.g., radio molecular line observations (e.g., Jijina et al. 1999; Caselli et al. 2002), dust emission/continuum observations (e.g., Kauffmann et al. 2008; Launhardt et al. 2010), Zeeman observations (e.g., Crutcher 1999; Crutcher et al. 2010), dust emission polarimetry (e.g., Ward-Thompson et al. 2000; Wolf et al. 2003), and dust dichroic extinction polarimetry (e.g., Jones et al. 2015; Kandori et al. 2017a,b). There is considerable observational data on the physical/chemical properties of cores, and important evidence has been reported (e.g., a tight geometrical relationship between the location of cores and filamentary structures, Andr\'{e} et al. 2010). However, obtaining direct observational constraints of the core formation process is extremely difficult. For example, there are no observational results for the initial radius $R_0$, initial density $\rho_0$, or initial magnetic field strength $B_0$ as the starting conditions of core formation. 
\par
To investigate the elementary process of core formation, we focused on the three-dimensional (3D) magnetic field structure of dense cores. Since the process must proceed from the accumulation of interstellar matter to create dense cores, and magnetic flux freezing is expected during the process, the most fundamental form of the magnetic field surrounding dense cores is expected to be hourglass shaped. The hourglass magnetic field is generated by core formation, and the history of mass condensation to create the core is reflected in the curvature of the hourglass field. Thus, a comparison of the appropriate flux freezing model with observations of the hourglass field can provide information on the initial conditions of core formation. 
\par
The object considered in the present study is the starless dense core FeSt 1-457. The fundamental physical parameters for FeSt 1-457 were determined based on density structure studies using the Bonnor--Ebert sphere model (Ebert 1955; Bonnor 1956). The radius, mass, and central density of the core are $R = 18500 \pm 1460$ AU ($144''$), $M_{\rm core} = 3.55 \pm 0.75$ $M_{\odot}$, and $\rho_{\rm c} = 3.5 (\pm 0.99) \times 10^5$ cm${}^{-3}$ (Kandori et al. 2005), respectively, at a distance of $130^{+24}_{-58}$ pc (Lombardi et al. 2006). The dimensionless radius parameter characterizing the Bonnor--Ebert density structure was $\xi_{\rm max} = 12.6 \pm 2.0$, which corresponds to a center to edge density contrast of $\rho_{\rm c} / \rho_{\rm s} = 75$. The subsequently measured background star polarimetry at near-infrared (NIR) wavelengths revealed an hourglass-shaped magnetic field toward the core (Kandori et al. 2017a, hereafter Paper I). Through simple modeling based on a 3D parabolic function, the structure of the 3D magnetic field (the magnetic field inclination angle toward the line of sight $\gamma_{\rm mag} = 35^{\circ} \pm 15^{\circ}$ and the 3D field curvature $C$) was determined (Kandori et al. 2017b, Paper II, see also, Appendix). Note that $\gamma_{\rm mag}$ is the line of sight inclination angle of the magnetic axis of the core measured from the plane of the sky. Since NIR polarization and extinction in FeSt 1-457 exhibit a linear relationship even in the dense region of the core, the above results reflect the overall dust alignment in the core (Kandori et al. 2018a, Paper III; Kandori et al. 2018c, Paper V). 
\par
From the $\gamma_{\rm mag}$ information, the total magnetic field strength of the core was determined to be $28.9 \pm 15.4$ $\mu$G using the Davis--Chandrasekhar--Fermi method (Davis 1951; Chandrasekhar \& Fermi 1953), which reveals the core to be in a magnetically supercritical state with $\lambda = 1.64 \pm 0.44$ (Paper II, see also, Appendix). Note that the total magnetic field strength at the core edge is $15$ $\mu$G, estimated based on an analysis of the magnetic field scaling on density (Kandori et al. 2018b, Paper IV; see also, Appendix). 
The value, $15$ $\mu$G, is consistent with the recently measured magnetic field strength for the inter-core regions of molecular clouds using the OH Zeeman effect ($\sim 15$ $\mu$G, Thompson et al. 2019). 
The stability of the core can be evaluated by comparing the observed mass of the core, $M_{\rm core}$, with the theoretical critical mass considering the magnetic and thermal/turbulent contributions in the core of $M_{\rm cr} \simeq M_{\rm mag} + M_{\rm BE}$ (Mouschovias \& Spitzer 1976; Tomisaka et al. 1988; McKee 1989), where $M_{\rm mag}$ is the magnetic critical mass and $M_{\rm BE}$ is the Bonnor--Ebert mass. The critical mass of the core is $M_{\rm cr} = 3.35 \pm 0.83$ $M_{\odot}$, which is comparable to the observed mass ($M_{\rm core} = 3.55 \pm 0.75$ $M_{\odot}$) of the core, suggesting that the core is in a nearly critical state. 
\par
In the present study, an analytic flux freezing magnetic field model (Myers et al. 2018, see also Mestel 1966; Ewertowski \& Basu 2013) was employed for comparison with the FeSt 1-457 data. The results were compared with our previous results (Paper II and Appendix) based on the axisymmetric parabolic function. The flux freezing model explained the FeSt 1-457 data well, and we derived the best-fit model parameters. With the obtained background density ($\rho_{\rm bkg}$) parameter and known core density, the initial contraction radius for core formation ($R_0$) and the initial magnetic field strength ($B_0$) were determined. Using these quantities, we discuss the initial conditions of the core formation and core formation mechanisms. 

\section{Data and Methods}
The NIR polarimetric data for FeSt 1-457 for the 3D magnetic field modeling was taken from Paper I. Observations were conducted using the $JHK_s$-simultaneous imaging camera SIRIUS (Nagayama et al. 2003) and its polarimetry mode SIRPOL (Kandori et al. 2006) on the IRSF 1.4-m telescope at the South African Astronomical Observatory (SAAO). SIRPOL can provide deep- (18.6 mag in the $H$ band, 5$\sigma$ in one hour exposure) and wide- ($7.\hspace{-3pt}'7 \times 7.\hspace{-3pt}'7$ with a scale of 0$.\hspace{-3pt}''$45 ${\rm pixel}^{-1}$) field NIR polarimetric data. 
\par
In the observed NIR polarimetric data, the polarization vectors toward FeSt 1-457 are superpositions of vectors arising from the core itself and from the core's ambient medium. The contribution from the ambient medium was removed in order to isolate the polarization vectors associated with the core (Paper I). 185 stars located within the core radius ($R \le 144''$) in the $H$ band were selected for the polarization analysis. Figure 1 shows the result. The magnetic field lines pervading the core have a shape reminiscent of an hourglass, which can be approximately traced using parabolic functions. 
\par
The existence of the distorted hourglass-shaped magnetic field can be interpreted as evidence for the mass condensation process. The curvature of the magnetic field lines in the outer region seems steep and the mass located outside the core should move across a large distance to create the current distorted magnetic field of the core. It is therefore clear that the core radius was previously larger than the current radius and that the core contracted by dragging the frozen-in magnetic field lines. 
\par
Since FeSt 1-457 is in a nearly kinematically critical state (Paper I; Paper II), the field distortion cannot be attributed to the dynamical collapse of the core. The observed distorted magnetic field is thus considered to be an imprint of the core formation process, in which mass was gathered and the magnetic field lines were dragged toward the center to create the dense core. These interpretations were presented in Paper I, and in the present study we quantitatively investigate core formation for FeSt 1-457 using a simple flux freezing model in an analytic form (Myers et al. 2018). 
\par
Examples of the distribution of the magnetic field lines using the flux freezing model (Myers et al. 2018) are shown in Figures 2 and 3. The model calculates the magnetic flux structures of spheroidal cores based on flux freezing and mass conservation. Since the projected shape of FeSt 1-457 is not elongated, we focus on the spherical case in the model. As initial conditions, we take a uniform magnetic field with a strength $B_0$ pervading the uniform medium with a density $\rho_0$. After the initiation of mass accumulation, isotropic contraction takes place, preserving the shape of the cloud during contraction. For the density structure, a Plummer-like model (Myers 2017) with an index $p=2$ was used. The index $p=2$ was chosen to approximate the density structure of the Bonnor--Ebert sphere (Ebert 1955; Bonnor 1956). The problem of mass loading in a flux tube was solved to connect the initially uniform density and flux distribution with the stage of mass and flux condensation arising from the cloud contraction. 
\par
In the model, the shape of the magnetic field lines, as shown in Figures 2 and 3, is a function of the density contrast $\rho_{\rm c} / \rho_{\rm bkg}$, where $\rho_{\rm c}$ is the density at the core center and $\rho_{\rm bkg}$ (alternatively, $\rho_0$) is the initial uniform density. Solutions with larger density contrast can result in a higher degree of central condensation in the magnetic field lines. 
\par
The equations in Myers et al. (2018) to obtain the magnetic field structure for a spherical core are as follows: 
\begin{eqnarray}
\xi_c &=& f_c^{1/2} \left [ 1+\frac{3 \nu_0}{\omega^2} \left( 1 - \frac{\tan^{-1} \omega}{\omega}  \right)  \right]^{-1/3}, \\
\zeta_c &=& (\omega^2 - \xi_c^2)^{1/2}.
\end{eqnarray}
Here, $\xi_c$ and $\zeta_c$ are dimensionless coordinates ($x$ and $z$ normalized to the scale length $r_0 \equiv \sigma / \sqrt{4 \pi G m n_0}$, where $\sigma$ is the one dimensional thermal velocity dispersion, $G$ is the gravitational constant, $m$ is the mean particle mass $2.33{\rm m_H}$, and $n_0$ is the peak density) representing the contours of the constant flux in the $xz$ plane (sky plane). $\nu_0 \equiv n_0 / n_u$ is the peak density normalized to the background value (density contrast). $\omega$ is the dimensionless radius of the sphere, which serves as a dummy variable increasing from $0$ to $\infty$. $f_c$ is the flux normalized to $\Phi_0 = \pi r_0^2 B_u$, where $B_u$ is the initial magnetic field strength.  
\par
Though the magnetic field structure of the flux freezing model looks similar to the structure derived using the parabolic model (2D: Paper I, 3D: Paper II and Appendix), they are not identical. Figure 4 shows a comparison between the magnetic field structure based on the flux freezing model (gray vectors, $\rho_{\rm c} / \rho_{\rm bkg} = 75$) and the parabolic fit to the flux freezing model data (black lines, $C=1.7 \times 10^{-6}$ pixel$^{-2}$ for the function $y=g+gCx^2$). Though the general trend in the structure of both model is the same, the gray vectors and black lines clearly deviate. Thus, we need to check whether the conclusions obtained using the parabolic model, especially in Paper II, can be reproduced for the flux freezing model. 
\par
The magnetic field structure shown in Figures 2 and 3 is the calculated result in the $xz$ plane (sky plane) of the spherical cloud core. To compare this with observations, we need to integrate the 3D polarization distribution toward the line of sight to derive the projected polarization map for various density contrast values. This process and the comparison with observations are described in the next section. 


\section{Results and Discussion}
\subsection{Application of Flux Freezing Model} 
3D polarization calculations of the flux freezing model (Myers et al. 2018) were made. 
Figures 2 and 3 show the calculation results on the $xz$ plane (sky plane). We assumed that the magnetic field lines are axisymmetric around the $z$ axis (radius $r$ and the direction $\phi$ around $z$ axis) in cylindrical coordinates. The model function $z(r,\phi,\rho_{\rm c} / \rho_{\rm bkg})$ thus has no dependence on the parameter $\phi$, where $\rho_{\rm c} / \rho_{\rm bkg}$ shows the density contrast for the core. 
For comparison with observations, after generating the model function, the 3D model is rotated in the line of sight ($\gamma_{\rm mag}$) and plane of sky ($\theta_{\rm mag}$) directions, and the axis of the cylindrical coordinates is set parallel to the direction of the magnetic axis (the orientation of the magnetic field pervading the core). The configuration of the coordinates and angles is shown in Figure 5. 
\par
For polarization modeling of the core, the 3D unit vectors of the polarization following the model function with a specific density contrast value were calculated using $750^3$ cells. Assuming that the orientation of the polarization vectors is parallel to the direction of the magnetic field, the 3D orientation of the polarization was determined in each cell. These unit vectors were then scaled to describe both the polarization angle and degree in each cell, $\Delta P_{H,{\rm model}}(x,y,z) = (\Delta P_{H,x}, \Delta P_{H,y}, \Delta P_{H,z})$. To determine the length of the polarization vector in each cell, we prepared the volume density value and the density--polarization conversion relationship. The volume density of molecular hydrogen in each cell, $n_{\rm H_2}(x,y,z)$, can be obtained from the known Bonnor--Ebert density structure of FeSt 1-457 ($\xi_{\rm max} = 12.6$, Kandori et al. 2005). The density--polarization conversion factor was estimated based on the slope of the $P_H$ vs. $H-K_s$ diagram of 4.8 \% mag$^{-1}$ (Paper I) as
\begin{equation}
|\Delta {\bf P}_{H,{\rm model}}(x,y,z)| = 0.22 \times w_{\rm cell} \times n_{\rm H_2}(x,y,z) / (9.4 \times 10^{20}),
\end{equation}
where $w_{\rm cell}$ is the size of the cell and $|\Delta {\bf P}|$ is the length of the $\Delta P$ vector in each cell. 
To obtain the scaling relationship, we used $A_V = 21.7 \times E_{H-K_s}$ (Nishiyama et al. 2008) and $N_{\rm H_2} / A_V = 9.4 \times 10^{20}$ cm$^{-2}$ mag$^{-1}$ (Bohlin, Savage, \& Drake 1978), where $N_{\rm H_2}$ is the column density of molecular hydrogen.
\par
The rotation of the polarization vector $\Delta P_{H,{\rm model}}(x,y,z)$ around the $x$-axis with an inclination angle $\gamma_{\rm mag}$ can be written as follows: 
\begin{eqnarray}
 \Delta P_{H,x}^{'} &=& \Delta P_{H,x}, \\
 \Delta P_{H,y}^{'} &=& \Delta P_{H,y} \cos\gamma_{\rm mag} - \Delta P_{H,z} \sin\gamma_{\rm mag}, \\
 \Delta P_{H,z}^{'} &=& \Delta P_{H,y} \sin\gamma_{\rm mag} + \Delta P_{H,z} \cos\gamma_{\rm mag}.
\end{eqnarray}
The data cube of $\Delta P_{H,{\rm model}}(x,y,z)$ is also rotated around the $x$ axis by an angle $\gamma_{\rm mag}$. 
\par
%
For sampling, $30^3$ cells were used, and the integrations of the cubes of the Stokes parameters toward the line of sight ($y$ direction) were conducted as 
\begin{eqnarray}
 q(x,z) &=& \int |\Delta P_{H,{\rm model}}^{'}(x,y,z)| \cos{2\theta_{\rm cell}} \cos^2{\gamma_{\rm cell}} \ dy, \\
 u(x,z) &=& \int |\Delta P_{H,{\rm model}}^{'}(x,y,z)| \sin{2\theta_{\rm cell}} \cos^2{\gamma_{\rm cell}} \ dy,
\end{eqnarray}
%
where $\theta_{\rm cell}(x,y,z)=\tan^{-1} (\Delta P_{H,z}^{'} / \Delta P_{H,x}^{'})$ is the position angle on the plane of sky and $\gamma_{\rm cell}(x,y,z)$ is the inclination angle with respect to the plane of sky in each cell. 
Since the magnetic field pervading the model core is distorted, the magnetic inclination angle in each cell $\gamma_{\rm cell}(x,y,z)$ is different from the inclination angle $\gamma_{\rm mag}$, which is the magnetic axis for the whole field. The $\gamma_{\rm cell}(x,y,z)$ angle can be calculated using the following equation: 
\begin{equation}
\gamma_{\rm cell}(x,y,z) = \cos^{-1}\sqrt{\frac{\Delta P_{H,x}^{'2} + \Delta P_{H,z}^{'2}}{|\Delta {\bf P^{'}}|^2}}.
\end{equation}
The polarization degree and angle can be obtained as
\begin{eqnarray}
 P_{H,{\rm model}}(x,z) &=& \sqrt{q^2(x,z) + u^2(x,z)}, \\
 \theta_{H,{\rm model}}(x,z) &=& \frac{1}{2}\tan^{-1}\left(  \frac{u(x,z)}{q(x,z)}  \right).
\end{eqnarray}
Finally, the orientation of the magnetic axis on the plane of sky, $\theta_{\rm mag} = 179^{\circ}$, was applied. $\theta_{H,{\rm model}}(x,z)$ was rotated by $\theta_{\rm mag}$ in both value and coordinates, and the $P_{H,{\rm model}}$ array was also rotated. 
\par
Figure 6 shows the polarization vector maps for the flux freezing model with a density contrast parameter of $\rho_{\rm c} / \rho_{\rm bkg} = 75$ for several line-of-sight inclination angles $\gamma_{\rm mag}$. In each panel of Figure 6, $\theta_{\rm mag}$ is set to $0^{\circ}$ for display. The white line shows the polarization vector, and the background color and color bar show the polarization degree of the model core. The applied viewing angle, $90^{\circ} - \gamma_{\rm mag}$, is labeled in the upper-left corner of each panel. Note that $90^{\circ} - \gamma_{\rm mag}$ is the angle between the direction toward the observer and the magnetic axis. 
\par
The features of the polarization vector maps in Figure 6 are similar to those in the 3D parabolic model described in Paper II, i.e., 1) a decrease of the maximum polarization degree from $\gamma_{\rm view} = 90^{\circ}$ to $\gamma_{\rm view} = 0^{\circ}$, 2) an hourglass-shaped polarization angle pattern that converges to a radial pattern toward small $\gamma_{\rm view}$, 3) depolarization in the polarization vector map, especially along the equatorial plane of the core, and 4) an elongated structure of the polarization degree distribution toward small $\gamma_{\rm view}$. 
\par
Figure 7 shows the $\chi^2$ distribution calculated using the model and observed polarization angle as 
\begin{equation}
\chi^2_{\theta} = \sum_{i=1}^{n}  \frac{(\theta_{{\rm obs},i} - \theta_{{\rm model},i})^2}{\delta \theta_i^2},
\end{equation}
where $n$ is the number of stars ($n=185$), $\theta_{{\rm obs},i}$ and $\theta_{{\rm model},i}$ denote the polarization angle from observations and the model for the $i$th star, and $\delta \theta_{{\rm obs},i}$ is the observational error. $\chi^2_\theta$ values were obtained for each inclination angle $\gamma_{\rm mag}$ after determining the best magnetic curvature parameter $C$. The inclination angle that minimizes $\chi^2_\theta$ is $\gamma_{\rm mag} = 35^{\circ}$, although the distribution of $\chi^2_\theta$ for the range between $\gamma_{\rm mag} = 0^{\circ}$ and $\sim 60^{\circ}$ is relatively flat. Note that the reduced $\chi^2$ values obtained in this analysis are large, because the relatively large variance originating from the Alfv\'{e}n wave cannot be included in the polarization angle error term, $\delta \theta_i^2$, in Equation (12). 
\par
Figure 8 shows the distribution of $\chi^2$ calculated using the model and observed polarization degree as 
\begin{equation}
\chi^2_{P} = \sum_{i=1}^{n}  \frac{(P_{{\rm obs},i} - P_{{\rm model},i})^2}{\delta P_{{\rm obs},i}^2},
\end{equation}
where $P_{{\rm obs},i}$ and $P_{{\rm model},i}$ represent the polarization degree from observations and the model for the $i$th star, and $\delta P_{{\rm obs},i}$ is the observational error. $\chi^2_P$ values were calculated for each $\gamma_{\rm mag}$ after minimizing the difference in polarization angles. 
\par
It should be noted here that the model polarization degree for each star $P_{{\rm model},i}$ was rescaled before calculating $\chi^2_P$. Though the scaling of $P_{{\rm model},i}$ was initially performed using Equation (3), it was without knowledge of the true magnetic inclination angle of the core. In other words, the factor in Equation (3) is the value assuming that the magnetic axis of the core is on the plane of sky. To correct this, we rescaled $P_{{\rm model},i}$ by the factor $\left< P_{\rm obs} / P_{\rm model} \right>$ determined using a robust least absolute deviation fitting. 
The mean values of $P_{{\rm model}}$ and $P_{{\rm obs}}$ are therefore always the same, and the deviation of the rescaled $P_{{\rm model}}$ from $P_{{\rm obs}}$ was calculated to evaluate $\chi^2_P$.  
%
\par
The minimization point for $\chi^2_P$ is the same inclination angle, $\gamma_{\rm mag} = 35^{\circ}$. We further conducted the same analysis using the 3D parabolic model (Appendix). The minimization angles, $\gamma_{\rm mag} = 35^{\circ}$ and $50^{\circ}$ were obtained for $\chi^2_{\theta}$ and $\chi^2_{P}$, respectively. On the basis of these analyses, we selected to use the value $\gamma_{\rm mag} = 35^{\circ} \pm 15^{\circ}$ throughout this paper.
%
\par
Figure 9 shows the relationship between $\chi^2_{\theta}$ and the density contrast $\rho_{\rm c} / \rho_{\rm bkg}$ when $\gamma_{\rm mag}$ is fixed to $35^{\circ}$. The minimization point of $\chi^2_{\theta}$ is $\rho_{\rm c} / \rho_{\rm bkg} \approx 85$. This is consistent with the value $\rho_{\rm c} / \rho_{\rm edge} \approx 75$ obtained based on the Bonnor--Ebert density profile analysis of FeSt 1-457 (Kandori et al. 2005). Two independent measurements, one based on the shape of the flux freezing magnetic field lines and the other based on the density profile, produce very consistent results. Hereafter we use a value of 75 for the density contrast of FeSt 1-457. 
\par
It is notable that the physical meaning of $\rho_{\rm edge}$ is different from that of $\rho_{\rm bkg}$. $\rho_{\rm edge}$ means the density at the core's boundary, which can be determined by comparing observations with the edge-truncated density profile model, such as the Bonnor--Ebert model. On the one hand, $\rho_{\rm bkg}$ means the initial density for the core formation or the diffuse uniform density at a large distance from core region, which can be determined by comparing the observed magnetic field structure of the core with the flux freezing magnetic field model. We found $\rho_{\rm bkg} \sim \rho_{\rm edge} \sim \rho_{\rm c} / 75 = 4670$ cm$^{-3}$ for FeSt 1-457. 
\par
Figure 10 shows the best-fit flux freezing model ($\gamma_{\rm mag} = 35^{\circ}$ and $\rho_{\rm c} / \rho_{\rm bkg} = 75$, white vectors) compared with observations (yellow vectors). The background image shows the distribution of the polarization degree. Figure 11 shows the same data but with the background image processed using the line integral convolution technique (LIC: Cabral \& Leedom 1993). We used the publicly available interactive data language (IDL) code developed by Diego Falceta-Gon\c{c}alves. The direction of the LIC \lq \lq texture'' is parallel to the direction of the magnetic field, and the background image is based on the polarization degree of the model core. 
The standard deviation of the polarization angle difference between the model and observations is $8.33^{\circ}$. This is comparable to the value $7.28^{\circ}$ for the 3D parabolic model case. 
\subsection{Core Formation of FeSt 1-457}
For an obtained core's density contrast, the initial density before core contraction ($\rho_0$) or the density of the inter-clump medium surrounding the core ($\rho_{\rm bkg}$) can be derived to be $\rho_{\rm c} / 75 = 4670$ cm$^{-3}$. This is about one order of magnitude higher than we expected for the inter-clump medium of the Pipe Nebula dark cloud complex. Radio molecular line observations toward the Pipe Nebula showed that 1) the overall distribution of $^{12}$CO ($J=1-0$) which traces $\sim 10^2$ cm$^{-3}$ gas is similar to that of the optical obscuration, and 2) the distribution of $^{13}$CO ($J=1-0$) which traces $\sim 10^3$ cm$^{-3}$ gas is similar to that of $^{12}$CO ($J=1-0$) (Onishi et al. 1999). The density of the overall diffuse inter-clump gas in the Pipe Nebula seems to be 10$^2$ to 10$^3$ cm$^{-3}$, while we expected a value of several $\times$ 10$^2$ cm$^{-3}$, in particular $\sim 300$ cm$^{-3}$ (Myers et al. 2018), for the density of inter-clump medium in the Pipe Nebula.
\par
The diffuse initial condition does not match the case for FeSt 1-457. If we assume this diffuse initial condition, the observed magnetic curvature should be steep, because in this case the magnetic curvature should follow the flux freezing model's solution of a density contrast one order of magnitude larger (see and compare, Figures 2 and 3). The solution of the model provides a steeper magnetic curvature as the density contrast increases. 
To explain the consistency between observations and the flux freezing model, FeSt 1-457 should be formed from the accumulation of relatively dense gas of several $\times$ 10$^3$ cm$^{-3}$. The core formation of FeSt 1-457 can be started from a relatively dense initial condition pervaded by a uniform magnetic field. 
In fact, FeSt 1-457 is located in a relatively dense region of the Pipe Nebula, in which the average $H-K_s$ color of stars is 0.4 mag in the reference field of FeSt 1-457 (Paper V), and $A_V \sim 5$ mag is expected in the Pipe Bowl region. The cloud thickness toward the Pipe Bowl region is $\sim 0.5$ pc (Franco et al. 2010). Dividing the background column density by the cloud thickness, we obtain $\sim 3000$ cm$^{-3}$ for the expected density for the Pipe Bowl region, which is comparable to the initial density ($\rho_0$) of FeSt 1-457 derived based on the magnetic field analysis. Thus, the suggestion of a relatively dense initial condition is observationally plausible. 
The {\it Herschel} observations of the Aquila Rift complex showed that $\sim 90$ \% of the candidate bound cores are found above a background dust extinction (column density) of $A_V \simgt 8$ mag (Andr\'{e} 2015, see also Onishi et al. 1998; Johnstone et al. 2004 for earlier ground-based studies). This is consistent with our scenario of relatively dense initial conditions for core formation. 
\par
The formation mechanism for such initial conditions is an open problem. A scenario of two colliding filamentary clouds in the Pipe Nebula region (Frau et al. 2015) may explain the relatively dense initial condition. The magnetic field can be compressed and can dominate in the Pipe Bowl region in the scenario involving the collision of filaments. The combination of the existence of a relatively dense inter-clump medium and a uniformly aligned magnetic field lines in the Pipe Nebula is not surprising. Alves et al. (2008) reported mass to flux ratio measurements of $\lambda_{\rm pos} \sim 0.4$ toward the Pipe Bowl region based on wide-field optical polarization observations. The existence of such a magnetically subcritical part is not special, because H\,{\sc i} clouds are known to be significantly magnetically subcritical (Heiles \& Troland 2005), and it is natural for molecular clouds, namely assemblies of diffuse H\,{\sc i} clouds, to have magnetically subcritical subregions. Since the magnetic field seems to dominate in the Pipe Bowl region, the field lines should be aligned even for the region of relatively high density. 
These results remind us of the classic ambipolar diffusion idea of slow drift of neutrals past nearly stationary field lines, followed by a more rapid supercritical collapse of an inner dense region (e.g., Mouschovias \& Ciolek 1999). In this scenario, the rapid collapse with flux freezing may be started at the density of several $\times 10^{3}$ cm$^{-3}$. Note that from Zeeman observations the density of $300$ cm$^{-3}$ was suggested as the point at which interstellar clouds become self-gravitating (Crutcher et al. 2010). 
\par
Since the initial density, $\rho_0$, is known through the analysis of the flux freezing model, the initial radius (core formation radius), $R_0$, can be obtained by $R_0 = (3 M_{\rm core} / 4 \pi \rho_{0})^{1/3}$, where $M_{\rm core}$ is the observed mass of the core. $R_0$ was calculated to be $1.64 R = 236'' = 0.15$ pc $= 30000$ au, where $R$ is the current radius of the core. In Figure 12, we show the extent of the core formation radius on the Digitized Sky Survey (DSS2, $R$ band) optical image of FeSt 1-457. The initial magnetic field strength $B_0$ was calculated to be $B_0 = B_{\rm tot} / 1.64^2 = 10.8$ $\mu$G, where $B_{\rm tot} = 28.9$ $\mu$G is the total magnetic field strength averaged for the whole core (Paper II and Appendix). It is notable that there are few methods available to obtain a dense core's initial radius ($R_0$), initial density ($\rho_0$), and initial magnetic field strength ($B_0$). 
\par
On the basis of obtained physical quantities, we consider the formation of FeSt 1-457. The Jeans mass $M_J$ of the core calculated using the initial density $\rho_0 = 4670$ cm$^{-3}$ is $4.21$ M$_{\odot}$ at 10 K. This value is consistent with the observed core mass of $M_{\rm core} = 3.55 \pm 0.75$ M$_{\odot}$. Moreover, the Jeans length is $\lambda_J = 0.32$ pc, which is close to the diameter of the core formation radius $2R_0 \approx 0.3$ pc. Though these results do not preclude the possibility of external compression by turbulence or shocks to create the core, the results of the Jeans analysis match the observations. The strength of gravity inside the formation radius of the core seems sufficient for initiating the formation of FeSt 1-457. 
\par
In addition to the Jeans analysis, we considered interstellar filaments for the origin of FeSt 1-457. In the non-magnetic case, an interstellar isothermal filament with gas temperature of 10 K has the critical mass per unit length $M_{\rm line, crit} = 2 c_{s}^2 / G \sim 16$ M$_{\odot}$ pc$^{-1}$ (Stod\'{o}lkiewicz 1963; Ostriker 1964; Inutsuka \& Miyama 1992). If we employ $R_0$ as a radius of the filament, the mean hydrogen molecule density of the critical filament is $4.3 \times 10^3$ cm$^{-3}$. In the magnetized case, following Tomisaka (2014), the critical mass per unit length can be $M_{\rm line, crit}^{\rm mag} \simeq 22.4 (R / 0.5\ {\rm pc}) (B / 10\ \mu{\rm G}) + 13.9 (c_s / 190\ {\rm m\ s}^{-1})$ M$_{\odot}$ pc$^{-1}$. We used $R_0$ as a radius of the filament and $B_0 = 10.4 - 14.6$ $\mu$G as a magnetic field strength in the filament (see the second last paragraph in this section for the estimation of $B_0$). The line mass and the mean hydrogen molecule density of the critical magnetized filament is $21-24$ M$_{\odot}$ pc$^{-1}$ and $5.7-6.4 \times 10^3$ cm$^{-3}$, respectively. These densities are well consistent with the initial density $\rho_0$ of FeSt 1-457. Therefore, the fragmentation of a filamentary cloud with nearly critical state can be the origin of FeSt 1-457. 
\par
Figure 13 shows the {\it Herschel} column density map (Roy et al. 2019; Andr\'{e} et al. 2010) covering the same spatial extent as Figure 12 ($30'$) around FeSt 1-457. The column density was converted to $A_V$ using $N_{\rm H_2}/A_V = 9.4 \times 10^{20}$ cm$^{-2}$ mag$^{-1}$ (Bohlin, Savage, \& Drake 1978). The resolution of the image is $18.2''$. In the map, there is a filamentary structure extending northward from FeSt 1-457, although the core seems relatively isolated especially toward the south. The Pipe Nebula dark cloud complex is well known for its filamentary shape, and the filamentary structure around FeSt 1-457 is small in scale compared with the global filament of the Pipe Nebula. Note that a network of sub-filaments within a large filament has been reported in the B59 region and the \lq \lq stem'' region in the Pipe Nebula (Peretto et al. 2012). 
\par
The mean density of the magnetized critical filament is slightly greater than $\rho_0$. The initial condition of the formation of FeSt 1-457 may be in slightly magnetically subcritical state. It is notable that the magnetized cylinder is unstable even when the magnetic field is extremely strong (Hanawa et al. 2017,2019). 
\par
The nearly critical filament was naturally derived from the analysis of the initial conditions of the formation of FeSt 1-457. This may be the result of supporting the \lq \lq interstellar filament paradigm'' (e.g., Andr\'{e} et al. 2014) from the core side. However, the initial diameter ($2R_0$) of FeSt 1-457 is $\sim 0.3$ pc, which is larger than the $0.1$ pc width obtained based on the {\it Herschel} data for a number of molecular clouds (e.g., Arzoumanian et al. 2011,2019). 
\par
A problem to employ this scenario is that there is no evidence of the infalling gas motion in FeSt 1-457 (Aguti et al. 2007). If the fragmentation of an interstellar filament can be the initial condition of core formation and the unstable condition evolves in a \lq \lq run-away'' fashion, the motion of gas moving inward of the core should be detected in observations, because FeSt 1-457 has been shrinking in radius from the initial radius $R_0=1.64R$ to the current radius $R$. 
\par
We speculate that the physical properties of the core born from the fragmentation of magnetically subcritical filament may be a key to explain the physical state of FeSt 1-457, because such a core can evolve in a quasi-static way until the mass to flux ratio of the core exceeds the critical value through the ambipolar diffusion. This scenario naturally explains rather static gas kinematics of FeSt 1-457. The model that best describes the structure of the core is the magnetohydrostatic model (e.g., Tomisaka et al. 1988). The stability of such configuration can be evaluated by the critical mass $M_{\rm cr} \simeq M_{\rm mag} + M_{\rm BE}$ (Mouschovias \& Spitzer 1976; Tomisaka et al. 1988; McKee 1989). $M_{\rm cr}$ decreases with decreasing magnetic critical mass $M_{\rm mag}$ through ambipolar diffusion, whereas there is a thermal support, which is represented in the equation by the Bonnor--Ebert mass $M_{\rm BE}$. Thus, if the thermal support is strong enough, the core can be stable even if the magnetic condition turns into supercritical. In this case, magnetically supercritical but quasi-static evolution continues until when the thermal and magnetic support is defeated by gravity. This scenario matches the physical conditions of FeSt 1-457, because the core is currently magnetically supercritical but kinematically nearly critical with additional support from the thermal pressure (Paper I, II, see also Appendix). 
\par
This scenario is also useful in explaining the hourglass structure of the magnetic field in FeSt 1-457. If the core is magnetically subcritical from birth to the present, the curvature of hourglass magnetic fields should be shallow, whereas the supercritical model can have more curvature in magnetic field lines (Basu et al. 2009). We expect that the most of the field curvature of FeSt 1-457 can be made during the magnetically supercritical phase of the core, and this should be investigated by comparing the observations of hourglass-like fields with theoretical simulations of dense core formation which include ambipolar diffusion process. 
%
\par
The free-fall time, $t_{\rm ff, ini}$, obtained based on the initial density $\rho_0$ of FeSt 1-457 is $\sim 5 \times 10^5$ yr. The sound crossing time, $t_{\rm sc, ini} \sim 1.5 \times 10^6$ yr, can be inferred from the initial core diameter $2R_0$ and nearly sonic internal velocity dispersion. These quantities, about one million years, serve as a lower limit value for the duration of starless phase of the core, and a factor of $\sim 2 - 6$ longer than the free-fall time calculated using the mean density of the current core ($t_{\rm ff, core} = 2.4 \times 10^5$ yr). The obtained factor, $\sim 2 - 6$, is consistent with the value $\sim 2 - 5$ (Ward-Thompson et al. 2007) estimated based on the number ratios of cores with and without embedded young stellar objects (e.g., Beichman et al. 1986; Lee \& Myers 1999; Jessop \& Ward-Thompson 2000). 
\par
It is known that the ambipolar diffusion timescale $t_{\rm AD}$ is about one order of magnitude longer than $t_{\rm ff}$ (e.g., McKee \& Ostriker 2007). The timescale of several times of $t_{\rm ff}$ is short for the evolution of the core with highly magnetically subcritical condition (e.g., Shu 1977). However, in a turbulent medium, the efficiency of ambipolar diffusion can be accelerated (e.g., Zweibel 2002; Fatuzzo \& Adams 2002; Nakamura \& Li 2005; Kudoh \& Basu 2014), and this may make $t_{\rm AD}$ reasonable length in timescale. 
Note that estimated starless time scale for FeSt 1-457 serves as lower limit, and it is still possible that FeSt 1-457 is a long-lived object. 
\par
The initial magnetic field strength $B_0$ is as weak as a typical inter-clump magnetic field in a molecular cloud (Crutcher 2012). The $B_0$ value was estimated by dividing the core's mean magnetic field strength $B_{\rm tot}$ by a geometrical dilution factor $1.64^2$. The actual initial magnetic field strength may be much larger, because the effect of ambipolar diffusion is not taken into account in the present calculation. 
The total magnetic field strength at the core boundary was estimated to be 14.6 $\mu$G (Paper IV, see also, Appendix). We thus consider the initial magnetic field strength $B_0$ to be in the range from $10.8$ to $14.6$ $\mu$G. Note that the value is consistent with the recently measured magnetic field strength for the inter-core regions of molecular clouds using the OH Zeeman effect ($\sim 15$ $\mu$G, Thompson et al. 2019). 
\par
Finally, we emphasize the importance of comparing observational (polarimetry) data with the theoretical flux freezing magnetic field model (e.g., Myers et al. 2018), with which we can obtain information on the initial conditions of core formation. A relatively dense initial condition may be common for core formation. Table 5 of Kandori et al. (2005) shows that the external pressure of dense cores is in the order of $10^4$ K cm$^{-3}$ based on Bonnor--Ebert density structure analyses. Assuming a gas temperature of 10 K, we find a relatively high value of $\sim 10^3$ cm$^{-3}$ for the density of the medium surrounding the dense cores, which is consistent with the case for FeSt 1-457 presented in this study. In order to determine common properties and regional property variations of dense cores, it is important to analyze a greater number of cores with the flux freezing magnetic field model. 

\section{Summary and Conclusion}
In the present study, the observational data for an hourglass-like magnetic field toward the starless dense core FeSt 1-457 were compared with a flux freezing magnetic field model (Myers et al. 2018). 
The flux freezing model gives a magnetic field structure consistent with observations. The best-fit parameters for the flux freezing model were a line-of-sight magnetic inclination angle of $\gamma_{\rm mag} = 35^{\circ}$ and a core center to ambient (background) density contrast of $\rho_{\rm c} / \rho_{\rm bkg} = 75$. Note that the same density contrast value was obtained through independent measurements based on a Bonnor--Ebert density structure analysis (Kandori et al. 2005). The initial density for core formation ($\rho_0$) was estimated to be $\rho_{\rm c} / 75 = 4670$ cm$^{-3}$, which is about one order of magnitude higher than the expected density ($\sim 300$ cm$^{-3}$) for the inter-clump medium of the Pipe Nebula. FeSt 1-457 is likely to have formed from the accumulation of relatively dense gas. The picture of a relatively dense initial condition for the formation of the core is supported by the relatively dense background column density ($A_V \simeq 5$ mag) around FeSt 1-457. The initial radius (core formation radius) $R_0$ and the initial magnetic field strength $B_0$ were obtained to be $1.64 R = 0.15$ pc and 10.8 $\mu$G, where $R$ is the current radius of the core. It is notable that there are few methods to obtain a dense core's initial physical parameters. The $B_0$ value is roughly consistent with a magnetic field strength measured at the core boundary of 14.6 $\mu$G (Paper IV). We thus conclude that the $B_0$ value is in the range from 10.8 to 14.6 $\mu$G. We found that the initial density $\rho_0$ is consistent with the mean density of the nearly critical magnetized filament with magnetic field strength $B_0$ and radius $R_0$. The relatively dense initial condition for core formation can be naturally understood if the origin of the core is the fragmentation of magnetized filaments. 

\subsection*{Acknowledgement}
We thank Takahiro Kudoh for helpful discussions. We are grateful to the staff of SAAO for their kind help during the observations. We with to thank Tetsuo Nishino, Chie Nagashima, and Noboru Ebizuka for their support in the development of SIRPOL, its calibration, and its stable operation with the IRSF telescope. The IRSF/SIRPOL project was initiated and supported by Nagoya University, National Astronomical Observatory of Japan, and the University of Tokyo in collaboration with the South African Astronomical Observatory under the financial support of Grants-in-Aid for Scientific Research on Priority Area (A) No. 10147207 and No. 10147214, and Grants-in-Aid No. 13573001 and No. 16340061 of the Ministry of Education, Culture, Sports, Science, and Technology of Japan. MT and RK acknowledge support by the Grants-in-Aid (Nos. 16077101, 16077204, 16340061, 21740147, 26800111, 19K03922). 

\subsection*{Appendix: Physical Properties of FeSt 1-457}
Here we summarize the physical properties of FeSt 1-457, measured by our group and others, for reference when referring to the series of FeSt 1-457 papers (Kandori et al. 2005, and Paper I, II, III, IV, V, and this paper). The FeSt 1-457 physical paramters are shown in Talbe S1 in A3. 
In addition, we report revised parameters and figures from the papers, especially Papers II and III (see A1) and V (see A2). The Stokes parameters ($q$ and $u$) determined through integration of the numerical cubes of the polarization parameters are shown in Equations (7) and (8) in Section 3.1. Though the same analysis was intended to be made in Paper II, the square in the $\cos^2 \gamma_{\rm cell}$ factor was absent in the calculations, and thus, we evaluated the effect of this and updated the physical parameters and figures. 
The line of sight inclination angle of the magnetic axis was revised from $\gamma_{\rm mag} = 45^{\circ} \pm 10^{\circ}$ (Paper II) to $35^{\circ} \pm 15^{\circ}$ (this paper). $\gamma_{\rm mag}$ is mainly used in the inclination correction of the physical parameters as the factor $1/\cos \gamma_{\rm mag}$, which changes by about $15$\% through the revision. Though this change is not large, it is not negligible. The revised figures from Paper II and III are presented in A1, and the revised parameters are shown in Table S1 in A3. In A1, the parameters derived using the parabolic magnetic field model are compared with the results based on the flux freezing model. 
%
In A2, we present the reanalyzed submillimeter polarimetry data (Alves et al. 2014,2015) of Paper V. The data was reanalyzed using a recently proposed method (Pattle et al. 2019), and the updated parameters are shown in Table S1 in A3. 
%
In A4, we compared our magnetic field strength measurements using the Davis-Chandrasekhar-Fermi method with the one based on the modified Davis-Chandrasekhar-Fermi method (Cho \& Yoo 2016; Yoon \& Cho 2019). 

\subsubsection*{A1: 3D Parabolic Model and Polarization--Extinction Relationship}
A 3D polarization calculation of the simple parabolic magnetic field model was conducted (Paper II and this paper). A 2D version of the model, $y=gCx^2$, was employed in Paper I, and we further assumed that the magnetic field lines are axisymmetric around the $z$ axis. The 3D function can be expressed as $z(r,\phi,g)=g+gCr^2$ in cylindrical coordinates $(r,z,\phi)$, where $g$ specifies the magnetic field line, $C$ is the curvature of the lines, and $\phi$ is the azimuth angle (measured on the plane perpendicular to $r$). This 3D function has no dependence on the parameter $\phi$. 
\par
After generating the model function, for comparison with observations, the 3D model is virtually observed after rotating in the line of sight ($\gamma_{\rm mag}$) and the plane of sky ($\theta_{\rm mag}$) directions. For this analysis, we followed the procedure described in Section 3.1 of this paper. The resulting polarization vector maps of the 3D parabolic model are shown in Figure S1. The white lines show the polarization vectors, and the background color and color bar show the polarization degree of the model core. The density structure of the model core was assumed to be the same as the Bonnor--Ebert sphere with a solution parameter of $12.6$ (the same parameter as obtained for FeSt 1-457, Kandori et al. 2005). The 3D magnetic curvature was set to $C = 2.0 \times 10^{-4}$ arcsec$^{-2}$ for all the panels. The applied viewing angle ($90^{\circ} - \gamma_{\rm mag}$), i.e., the angle between the line of sight and the magnetic axis, is labeled in the upper left corner of each panel. 
\par
The model polarization vector maps change depending on the viewing angle ($\gamma_{\rm view}$). As described in Paper II, there are four characteristics: 1) a decrease of maximum polarization degree from $\gamma_{\rm view}=90^{\circ}$ to $\gamma_{\rm view}=0^{\circ}$ , 2) an hourglass-shaped polarization angle pattern for large $\gamma_{\rm view}$ converges to a radial pattern for small $\gamma_{\rm view}$, 3) depolarization occurs in the polarization vector map, especially along the equatorial plane of the core, and 4) an elongated structure of the polarization degree distribution toward small $\gamma_{\rm view}$. Compared with the case of the flux freezing model (Figure 6), there are some differences in Figure S1, especially for the low $\gamma_{\rm view}$ regions. However, both models have the above four characteristics, showing a similar dependence of the polarization features on $\gamma_{\rm view}$. For details of these characteristics, see Section 3.1 of Paper II. 
\par
Figures S2 and S3 show the $\chi^2$ distributions with respect to the polarization angle and degree ($\chi^2_{\theta}$ and $\chi^2_P$). The calculation methods are the same as those described in Section 3.1, and the minimization points are $35^{\circ}$ for $\chi^2_{\theta}$ and $50^{\circ}$ for $\chi^2_{P}$. Since we obtained $35^{\circ}$ for both $\chi^2_{\theta}$ and $\chi^2_{P}$ using the flux freezing model in Section 3.1, we concluded that the line of sight inclination angle $\gamma_{\rm mag}$ is $35^{\circ} \pm 15^{\circ}$. 
\par
Figure S4 shows the best-fit 3D parabolic model ($\gamma_{\rm mag}=35^{\circ}$ and $C=2.0 \times 10^{-4}$ arcsec$^{-2}$, white vectors) compared with observations (yellow vectors). The background image shows the distribution of the polarization degree. Figure S5 shows the same data but with the background image processed using the line integral convolution technique (LIC: Cabral \& Leedom 1993). The direction of the LIC \lq \lq texture'' is parallel to the direction of the magnetic field, and the background image is based on the polarization degree of the model core. The results look similar to the flux freezing model case (Figures 10 and 11). 
\par
Figures S6--S9 show the polarization--extinction ($P$--$A$) relationship measured at NIR wavelengths. The linearity in the $P$--$A$ relationship is important in two respects: it shows that the observed polarization vectors trace the magnetic field structure inside the core, and it can be used to compare the relationship with theories of dust grain alignment (e.g., grain alignment with radiative torque: Dolginov \& Mitrofanov 1976; Draine \& Weingartner 1996, 1997; Lazarian \& Hoang 2007). 
Comparing Figure S6 with Figure 4 of Paper III, panels (a) and (b) are the same, and the shapes of the plots in panels (c) are very similar except for the slope. Note that in panel (c) we corrected the effects of depolarization and the line-of-sight inclination at the same time by dividing the panel (b) relationship by the 2D array of correction factors (Figure S9), so that panel (c) corresponds to panel (d) in Figure 4 of Paper III. In the revision, the $\cos^2 \gamma_{\rm mag}$ factor with the angle $35^{\circ}$ was used in the calculations for panel (c). This does not change the linearity of the plot but changes the steepness in the slope. The slope, $P_H/E_{H-K_s}$, for each panel is $2.43 \pm 0.05$, $4.76 \pm 0.33$, and $6.60 \pm 0.41$ \% mag$^{-1}$ for the panel (a), (b), and (c), respectively. 
%
%
Figure 4(b) of Paper V was revised in the same way and the corrected relationships are shown in Figures S7 and S8. The dotted line in Figures S7 and S8 shows the power-law fitting to the data, resulting in $\alpha_H = -0.07 \pm 0.11$ for the relationship $P_H / A_V \propto A_V^{-\alpha_H}$. The dashed line in Figure S7 shows the linear fitting to the data, resulting in a slope of $0.002 \pm 0.002$ \% mag$^{-1}$. The dotted-dashed lines in Figures S7 and S8 show the observational upper limit as determined by Jones (1989). The relation was calculated based on the equation $P_{K,{\rm max}} = \tanh \tau_{\rm p}$, where $\tau_{\rm p} = (1-\eta)\tau_K / (1+\eta)$, and the parameter $\eta$ is set to 0.875 (Jones 1989). $\tau_K$ denotes the optical depth in the $K$ band, and $P_H / A_V \approx 0.62$ at $\tau_K = 1$. 
Note that though the above revisions are minor in terms of the shape/linearity of the plots, the steepness of the slope is important when we discuss the efficiency of dust grain alignment. 
\par
The correlation coefficients for the Figure S6 relationship are 0.68, 0.76, and 0.85 for the panels (a), (b), and (c), respectively. It is evident that the corrections (subtraction of ambient off-core polarization components, depolarization correction, and inclination correction) improve the tightness in the polarization--extinction relationship. The obtained $P_H / A_V$ versus $A_V$ relationship shows a flat distribution. The $\alpha_H$ index for $P_H / A_V \propto A_V^{-{\alpha}_H}$ is negative, although the value is consistent with $\alpha_H = 0$. This indicates that the magnetic field pervading FeSt 1-457 is fairly uniform, at least for the range probed in the present observations ($A_V \simlt 25$ mag). It is also clear that our NIR polarimetric observations trace the polarizations arisen inside the core. 
\par
Finally, we explain Figure S9, showing the depolarization and inclination correction factor. To obtain the factor, we divided the $\gamma_{\rm mag} = 35^{\circ}$ model by the $\gamma_{\rm mag} = 0^{\circ}$ model with the same magnetic curvature. In Figure S9, the factors in the regions around the equatorial plane are less than unity, showing that the depolarization effect applies. This is due to the crossing of the polarization vectors at the front and back sides of the core along the line of sight (see the explanatory illustration of Figure 7 of Kataoka et al. 2012). In the upper and lower regions of the map, the factors have values around unity. While we would expect a value of $\cos^2 \gamma_{\rm mag} = \cos^2 35^{\circ} = 0.67$ for the case of a uniform field, for the parabolic field case, most of the magnetic field lines around the poles are inclined with respect to the magnetic axis, reducing the polarization degree in the regions in the $\gamma_{\rm mag} = 0^{\circ}$ model and consequently increasing the correction factors from $0.67$. 
%
%

\subsubsection*{A2: Power-law Index of Submillimeter Polarimetry Data}
As shown in Figures S7 and S8, the polarization efficiency at NIR wavelengths is nearly constant against $A_V$, indicating that the observations trace the dust alignment, i.e., the magnetic field structure, in FeSt 1-457. However, the probing depth in our polarimetry is limited to $A_V \sim 25$ mag. To investigate the magnetic field structure deep inside the core, polarimetric observations at longer wavelengths are important. In Paper V, using the data of Alves et al. (2014, 2015), obtained with the APEX 12-m telescope and PolKa polarimeter at $870$ $\mu$m (for the instrument see Siringo et al. 2004, 2012; Wiesemeyer et al. 2014), we showed that the magnetic field orientations obtained from submillimeter polarimetry ($132.1^{\circ} \pm 22.0^{\circ}$) and NIR polarimetry ($2.7^{\circ} \pm 16.2^{\circ}$) differ significantly. This may indicate a change of magnetic field orientation inside the core. However, the polarization fraction at submillimeter wavelengths $P_{\rm submm}$ has an $\alpha_{\rm submm}$ index of $0.92 \pm 0.17$ for the $P_{\rm submm} \propto I_{\rm submm}^{-\alpha_{\rm submm}}$ relationship (Alves et al. 2015). An $\alpha_{\rm submm}$ index close to unity indicates that the alignment of dust inside the core should be lost (e.g., Andersson et al. 2015). 
\par
The polarization fraction data points obtained with dust emission polarimetry are usually debiased (e.g., Wardle \& Kronberg 1974) and points having a signal to noise ratio (SNR) larger than a certain value are selected for the power-law fitting. Recently, Pattle et al. (2019) reported that the usual method for obtaining the $\alpha$ power-law index can lead to an overestimation of $\alpha$, and demonstrated that the Ricean-mean model fitting to the whole data (without debias) can provide a better estimation of the $\alpha$ index. We followed this method to revise/improve the $\alpha_{\rm submm}$ index. The $P_{\rm submm}$ versus $I_{\rm submm}$ data were fitted using the following equation: 
\begin{equation}
P_{\rm submm}= \sqrt{\frac{\pi}{2}} \left( \frac{I_{\rm submm}}{\sigma_{QU}} \right)^{-1} \mathcal{L}_{\frac{1}{2}} \left [ -\frac{P^2_{\sigma_{QU}}}{2} \left(  \frac{I_{\rm submm}}{\sigma_{QU}}   \right)^{2(1-\alpha_{\rm submm})}  \right].
\end{equation}
This is taken from Equation (21) in Pattle et al. (2019), which they refer to as the Riceal-mean model. In the equation, $\sigma_{QU}$ is the RMS noise in the Stokes $Q$ and $U$ measurements, $P_{\sigma_{QU}}$ is a parameter to be fitted simultaneously with $\alpha_{\rm submm}$, and $\mathcal{L}_{\frac{1}{2}}$ is a Laguerre polynomial of order $\frac{1}{2}$. We fitted the observations using this function, and the results are shown in Figure S10 as a solid line. The dotted line shows the relationship for the low-SNR limit defined by Equation (12) of Pattle et al. (2019). Note that the $P_{\rm submm}$ values greater than unity are physically meaningless. 
The best-fit parameters are $\alpha_{\rm submm} = 0.41 \pm 0.10$ and $P_{\sigma_{QU}} = 0.30 \pm 0.10$. We obtained a significantly low value of $\alpha_{\rm submm}$ compared with the fitting based on the ordinary method (Alves et al. 2015). Thus, we conclude that the alignment of dust grains is better than previously thought. 

\subsubsection*{A3: List of Physical Parameters}
In Table S1, we summarize the physical parameters for FeSt 1-457. This parameter list does not contain all the values reported so far, but shows the physical parameters mainly used in our studies related to this core (Kandori et al. 2005; Paper I, II, III, IV, V, and this paper). For example, the parameters for the chemical properties reported by Juarez et al. (2017) or the dust grain (growth) properties reported by Forbrich et al (2015) are not included. 

\subsubsection*{A4: Modified Davis-Chandrasekhar-Fermi method} 
Cho \& Yoo (2016) and Yoon \& Cho (2019) studied the reduction of variation in polarization angle $\delta \theta$ due to the averaging effect along the line of sight. If there is more than one independent turbulent eddy along the line of sight, the measured value of $\delta \theta$ will be reduced. They suggested to use $\delta V_{\rm c}$, the standard deviation of centroid velocity of optically thin molecular line, instead of $\sigma_{\rm turb}$, the turbulent velocity dispersion, in the original Davis-Chandrasekhar-Fermi formulation. The conventional form of Davis-Chandrasekhar-Fermi method is 
\begin{equation}
B = C_{\rm corr} \sqrt{4 \pi \bar{\rho}} \frac{\sigma_{\rm turb}}{\delta \theta}, 
\end{equation}
where $\bar{\rho}$ is mean density, and $C_{\rm corr} = 0.5$ is a correction factor suggested by theoretical studies (Ostriker et al. 2001, see also, Padoan et al. 2001; Heitsch et al. 2001; Heitsch et al. 2005; Matsumoto et al. 2006). The modified Davis-Chandrasekhar-Fermi method is 
\begin{equation}
B = \xi \sqrt{4 \pi \bar{\rho}} \frac{\delta V_{\rm c}}{\delta \theta}, 
\end{equation}
where $\xi$ is a constant of order unity that can be determined by numerical simulations. The standard deviation of centroid velocity is given by
\begin{equation}
\delta V_{\rm c} \approx \frac{\sigma_{\rm turb}}{\sqrt{N_{\rm eddy}}}
\end{equation}
where $N_{\rm eddy}$ is the number of independent turbulent eddy along the line of sight. 
\par
We obtained $\sigma_{\rm turb} = 0.0573 \pm 0.006$ km s$^{-1}$ based on the N$_2$H$^+$ ($J=1-0$) molecular line observations using the Nobeyama 45m radio telescope (Kandori et al. 2005). Using the same data, we obtained $\delta V_{\rm c} \approx 0.023$ km s$^{-1}$. Note that the standard deviation of $V_{\rm c}$ was calculated after subtracting the rigid rotation component estimated by plane fitting. Comparing this value with $C_{\rm corr} \times 0.0573 = 0.029$ km s$^{-1}$, the difference is about 20\%, indicating that the applications of Davis-Chandrasekhar-Fermi method and its modified version to FeSt 1-457 yield consistent results. The expected number of independent turbulent eddy is $\approx 6.2$. The relatively small $N_{\rm eddy}$ enables the use of classic Davis-Chandrasekhar-Fermi formula for FeSt 1-457, and such situations might be common for other low mass dense cores.

\clearpage 

\begin{figure}[t]
\begin{center}
 \includegraphics[width=6.5 in]{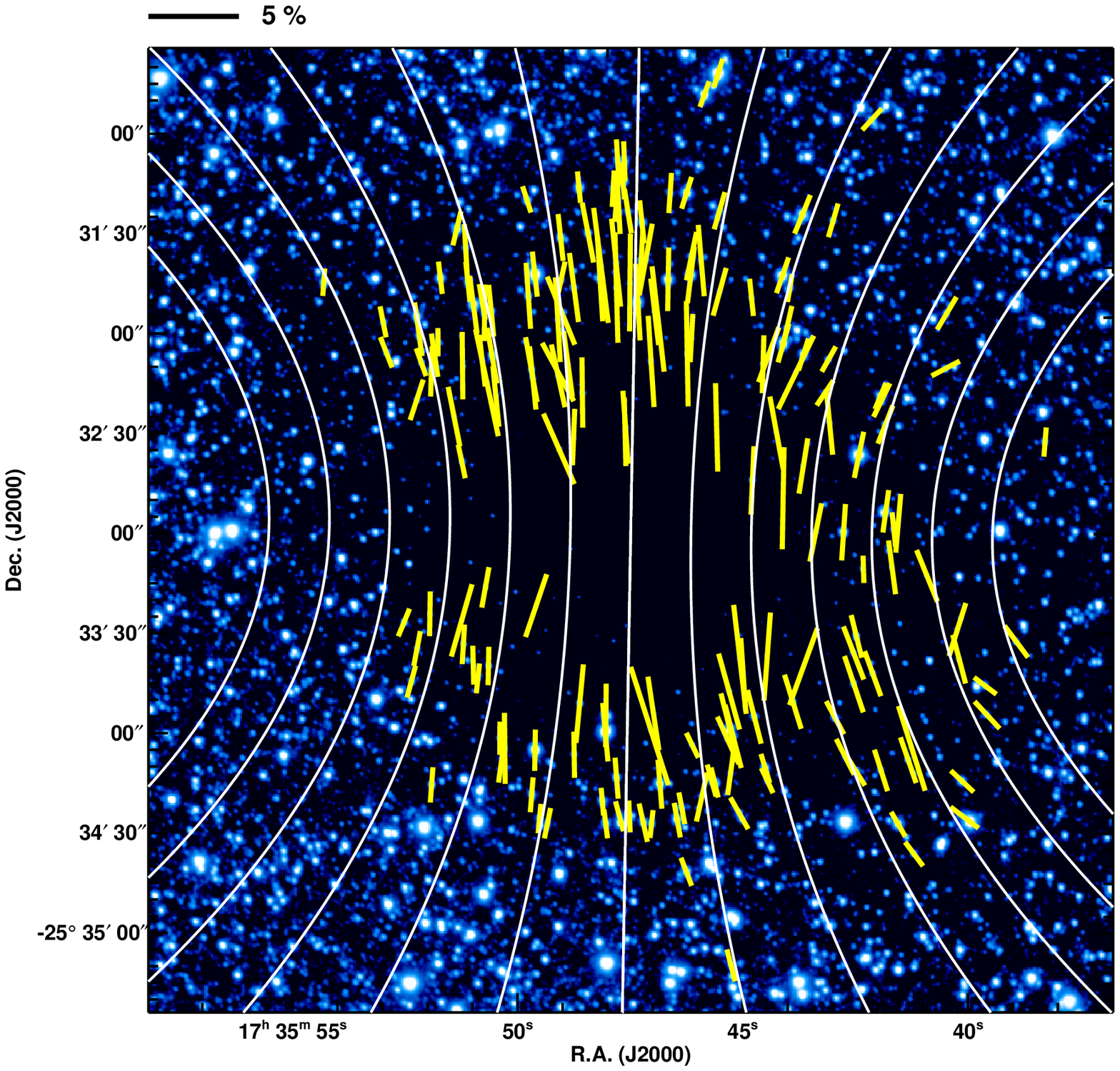}
\end{center}
 \caption{Polarization vectors for FeSt 1-457 after subtraction of the ambient polarization component (yellow vectors). The figure is taken from Paper I. The field of view is the same as the diameter of the core ($288'' = 0.19$ pc). The white lines show the magnetic field direction inferred from fitting with a parabolic function $y=g+gCx^2$, where $g$ specifies the magnetic field lines and $C$ determines the degree of curvature in the parabolic function. The scale of 5\% polarization degree is shown at the top.}
   \label{fig1}
\end{figure}

\clearpage 

\begin{figure}[t]
\begin{center}
 \includegraphics[width=6.5 in]{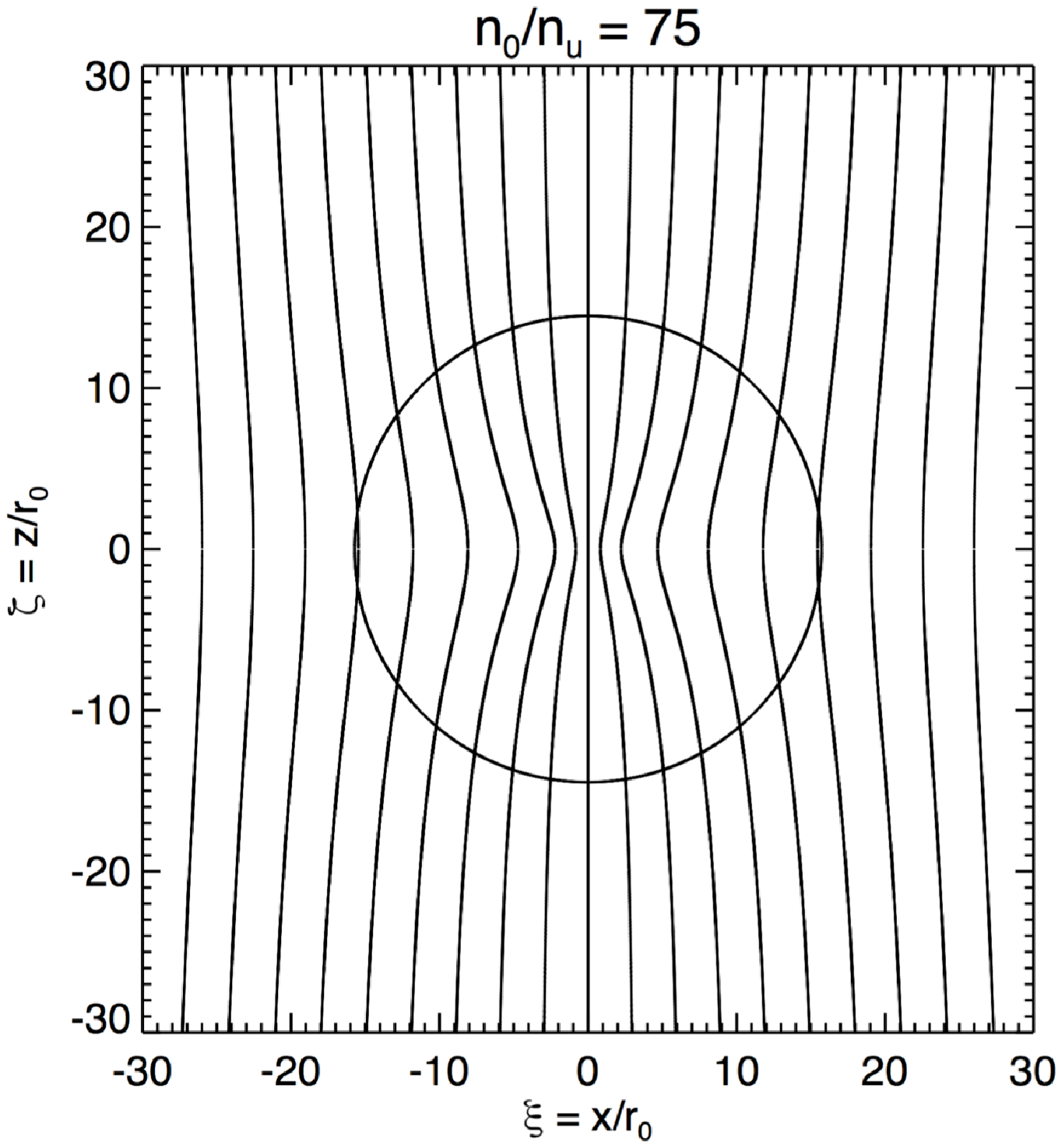}
\end{center}
 \caption{Distribution of magnetic flux contours based on the flux freezing model by Myers et al. (2018). Results for a density contrast parameter of 75 are shown. The circle shows the core radius. The $xz$ plane of the core is shown, and both the $x$ and $z$ axes are normalized by the scale length $r_0 \equiv \sigma / \sqrt{4 \pi G \rho_0}$, where $\sigma$ is the one-dimensional thermal velocity dispersion, $G$ is the gravitational constant, and $\rho_0$ is the background density.}
   \label{fig1}
\end{figure}

\clearpage 

\begin{figure}[t]
\begin{center}
 \includegraphics[width=6.5 in]{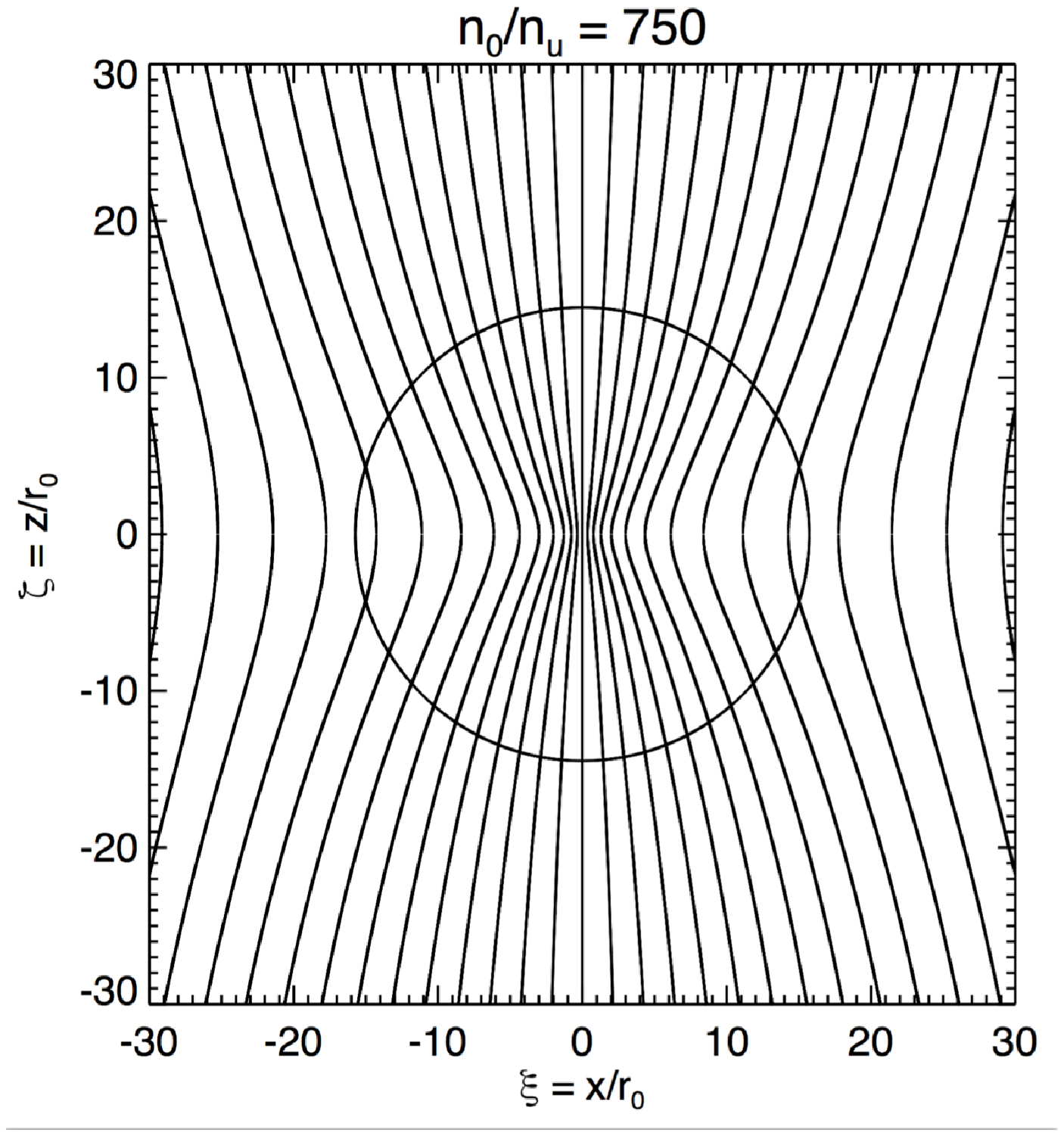}
\end{center}
 \caption{Distribution of magnetic flux contour based on the flux freezing model by Myers et al. (2018). Results for a density contrast parameter of 750 are shown. The circle shows the core radius. The $xz$ plane of the core is shown, and both the $x$ and $z$ axes are normalized by the scale length $r_0 \equiv \sigma / \sqrt{4 \pi G \rho_0}$, where $\sigma$ is the one-dimensional thermal velocity dispersion, $G$ is the gravitational constant, and $\rho_0$ is the background density.}
   \label{fig1}
\end{figure}

\clearpage 

\begin{figure}[t]
\begin{center}
 \includegraphics[width=6.5 in]{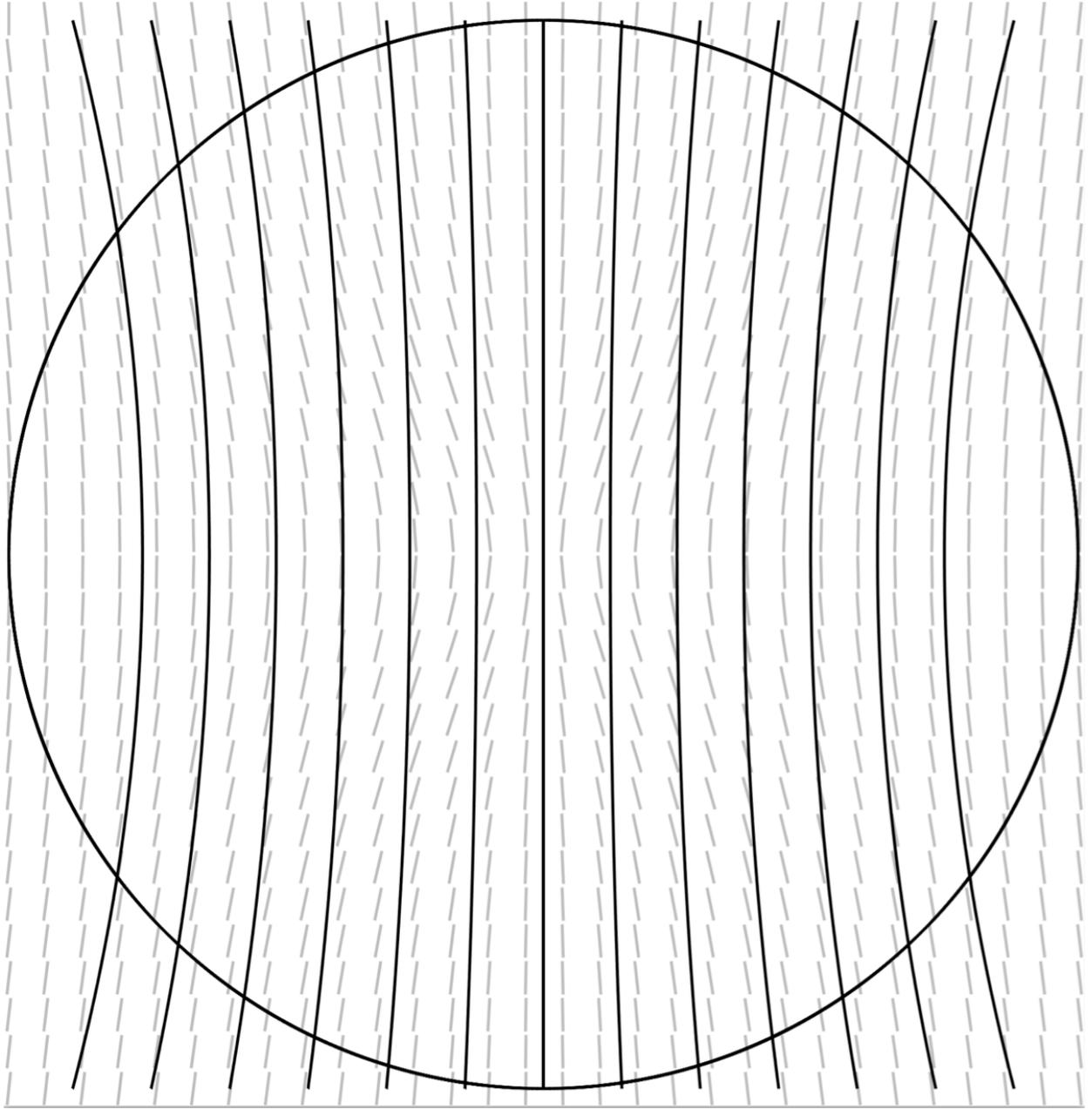}
\end{center}
 \caption{Comparison of magnetic field structures based on the flux freezing model ($\rho_{\rm c}/\rho_{\rm bkg} = 75$) and the parabolic model ($C=1.7 \times 10^{-6}$ pixel$^{-2}$ for the function $y=g+gCx^2$). The comparison was done on the $xz$ plane. The circle shows the radius of the core.}
   \label{fig1}
\end{figure}

\clearpage 

\begin{figure}[t]
\begin{center}
 \includegraphics[width=6.5 in]{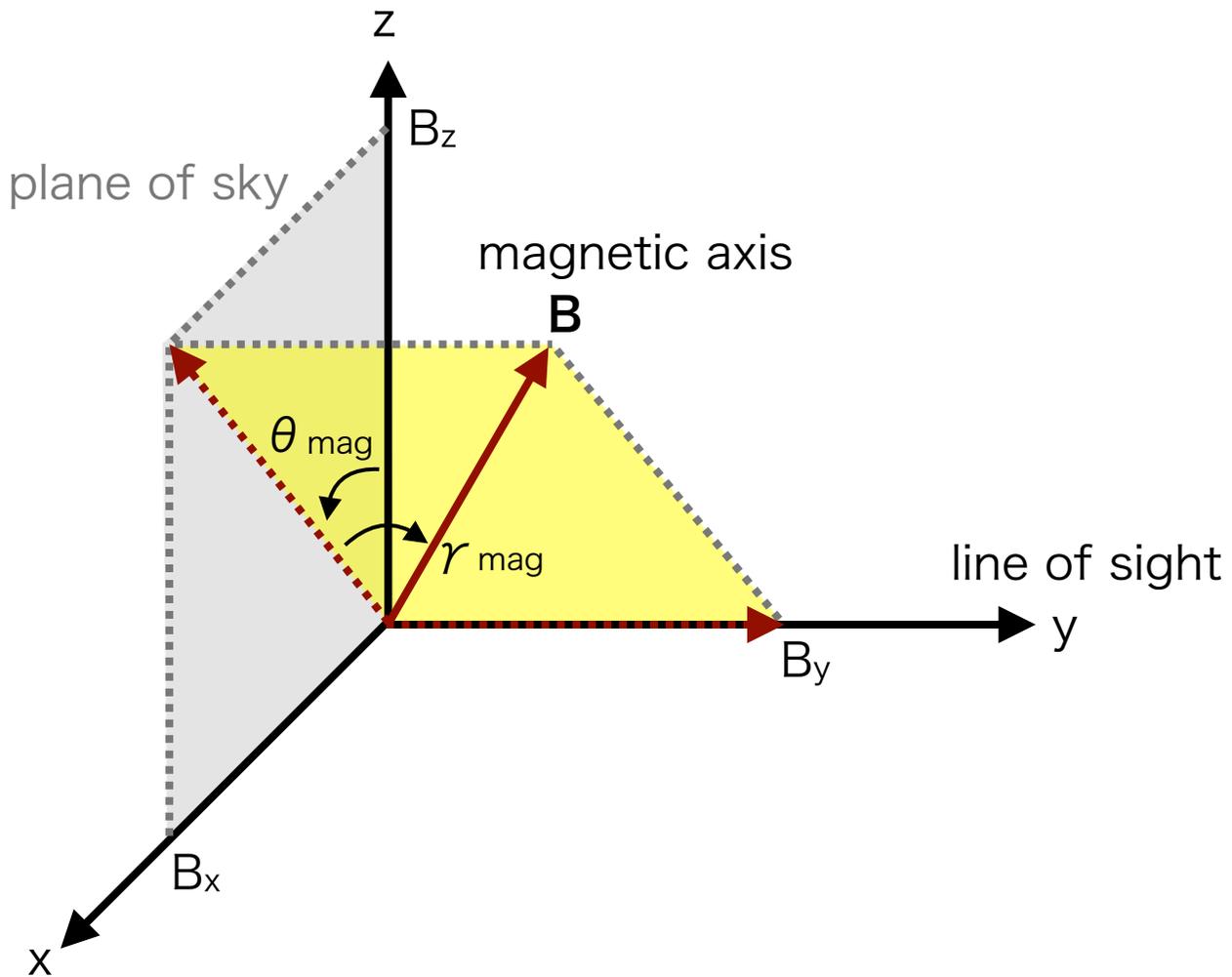}
\end{center}
 \caption{Configurations of coordinates and angles. The $z$ axis is toward the zenith, and the $xz$ plane corresponds to the plane of sky. The $y$ direction is toward the line of sight. $\gamma_{\rm mag}$ and $\theta_{\rm mag}$ show the line of sight and plane of sky inclination angles of the magnetic axis, respectively.}
   \label{fig1}
\end{figure}

\clearpage 

\begin{figure}[t]
\begin{center}
 \includegraphics[width=6.5 in]{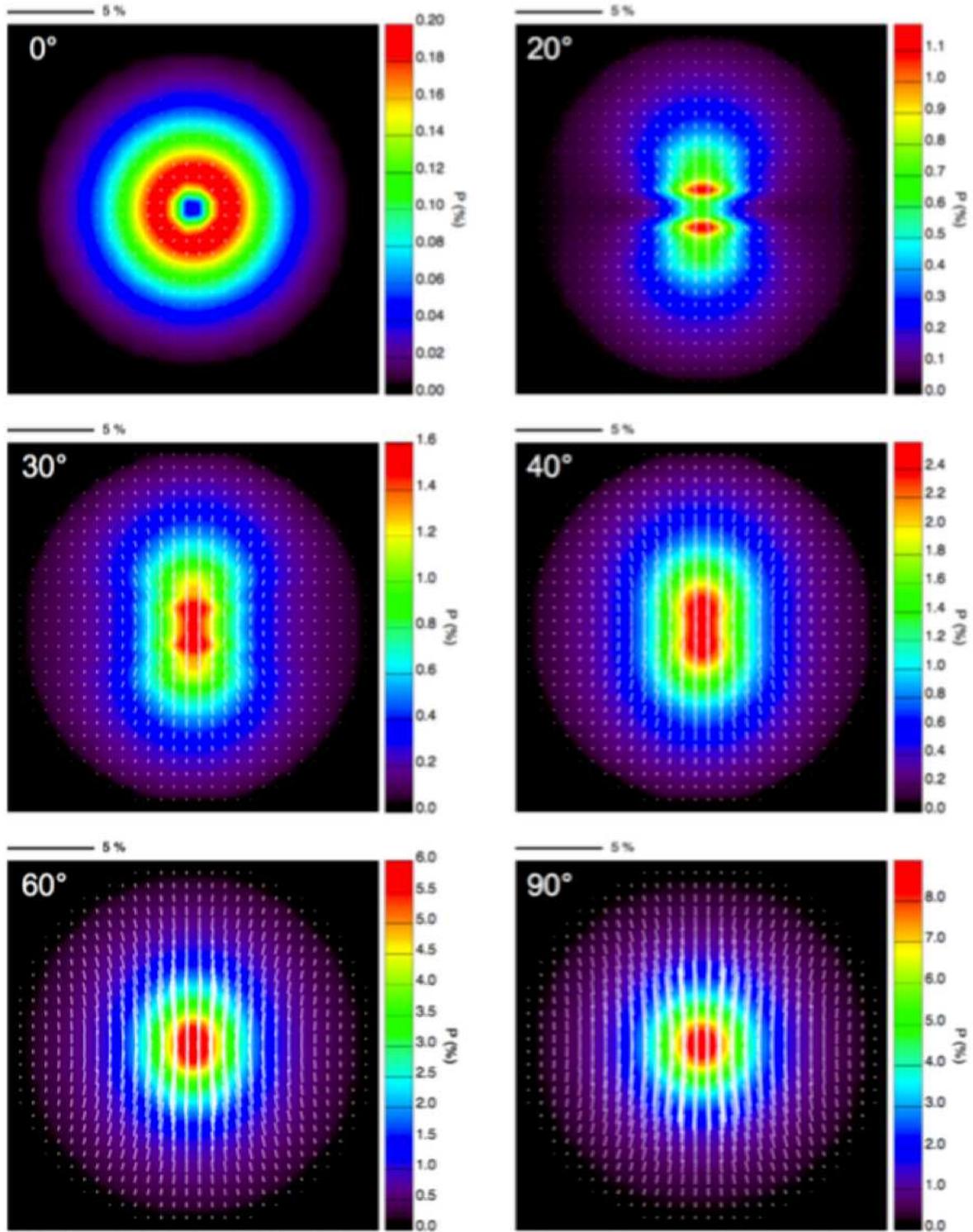}
\end{center}
 \caption{Polarization vector maps of the 3D flux freezing model (white vectors). The background color and color bar show the polarization degree. The applied viewing angle ($\gamma_{\rm view} = {90}^{\circ} - \gamma_{\rm mag}$) is labeled in the upper left corner of each panel. The density contrast parameter ($\rho_{\rm c} / \rho_{\rm bkg}$) is set to 75 for all the panels.}
   \label{fig1}
\end{figure}

\clearpage 

\begin{figure}[t]
\begin{center}
 \includegraphics[width=6.5 in]{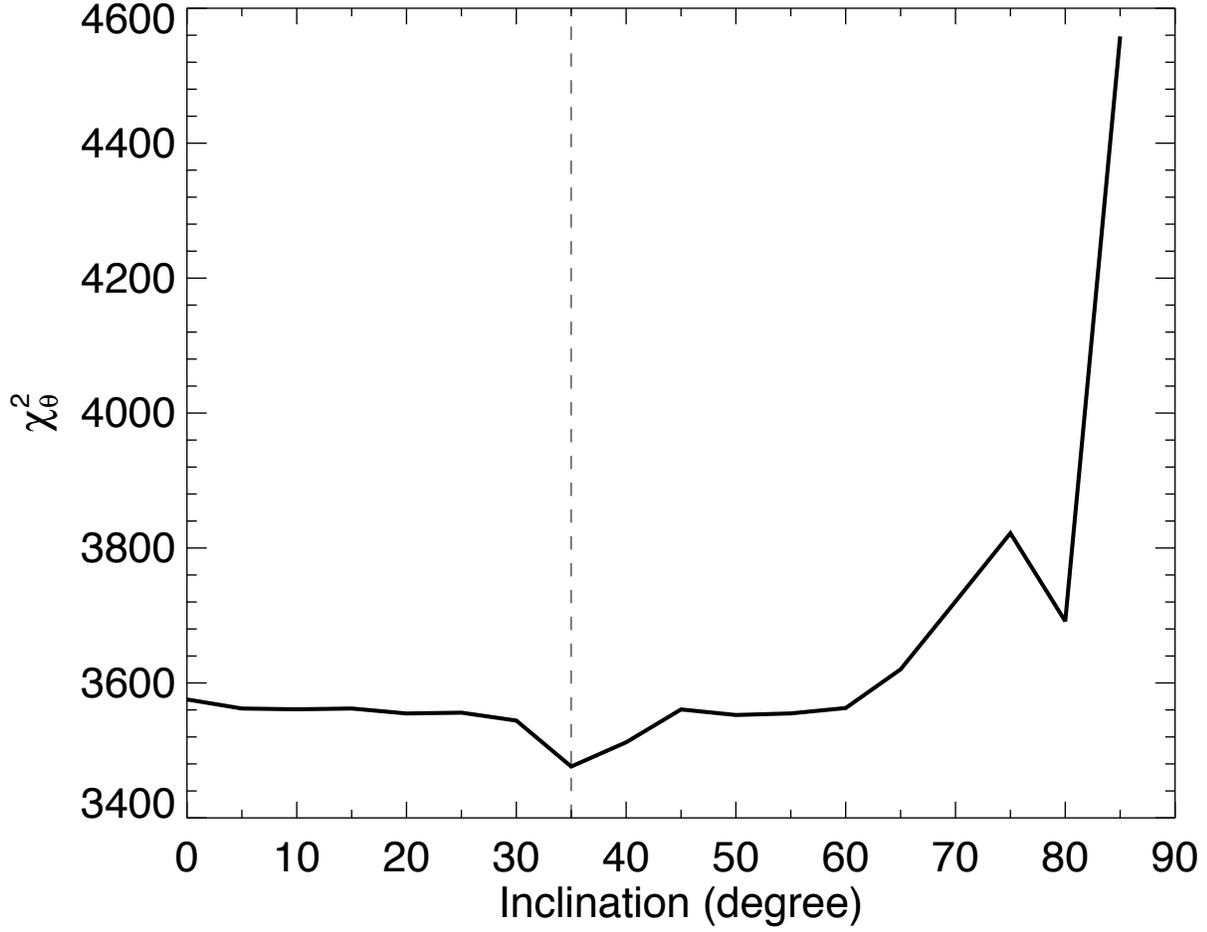}
\end{center}
 \caption{$\chi^2$ distribution of the polarization angle ($\chi_{\theta}^{2}$). The best density contrast parameter ($\rho_{\rm c} / \rho_{\rm bkg}$) was determined for each inclination angle ($\gamma_{\rm mag}$). $\gamma_{\rm mag} = 0^{\circ}$ and $90^{\circ}$ correspond to the edge-on and pole-on geometries with respect to the magnetic axis.}
   \label{fig1}
\end{figure}

\clearpage 

\begin{figure}[t]
\begin{center}
 \includegraphics[width=6.5 in]{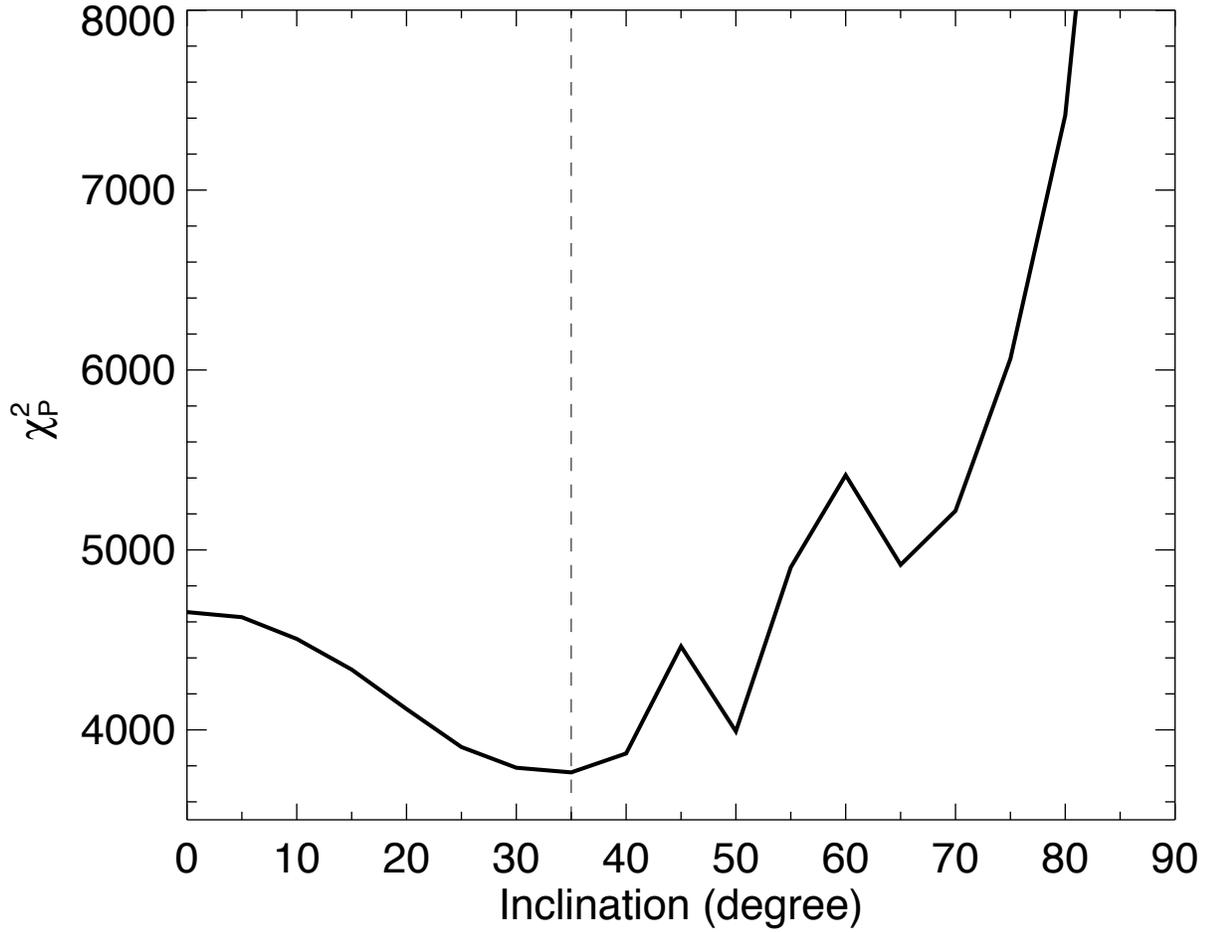}
\end{center}
 \caption{$\chi^2$ distribution of the polarization degree ($\chi_{P}^{2}$). The calculations of $\chi^2$ in polarization degree were performed after determining the best density contrast parameter ($\rho_{\rm c} / \rho_{\rm bkg}$) that minimizes $\chi^2$ in the polarization angle. This calculation was carried out for each $\gamma_{\rm mag}$. $\gamma_{\rm mag} = 0^{\circ}$ and $90^{\circ}$ correspond to the edge-on and pole-on geometries in the magnetic axis.}
   \label{fig1}
\end{figure}

\clearpage 

\begin{figure}[t]
\begin{center}
 \includegraphics[width=6.5 in]{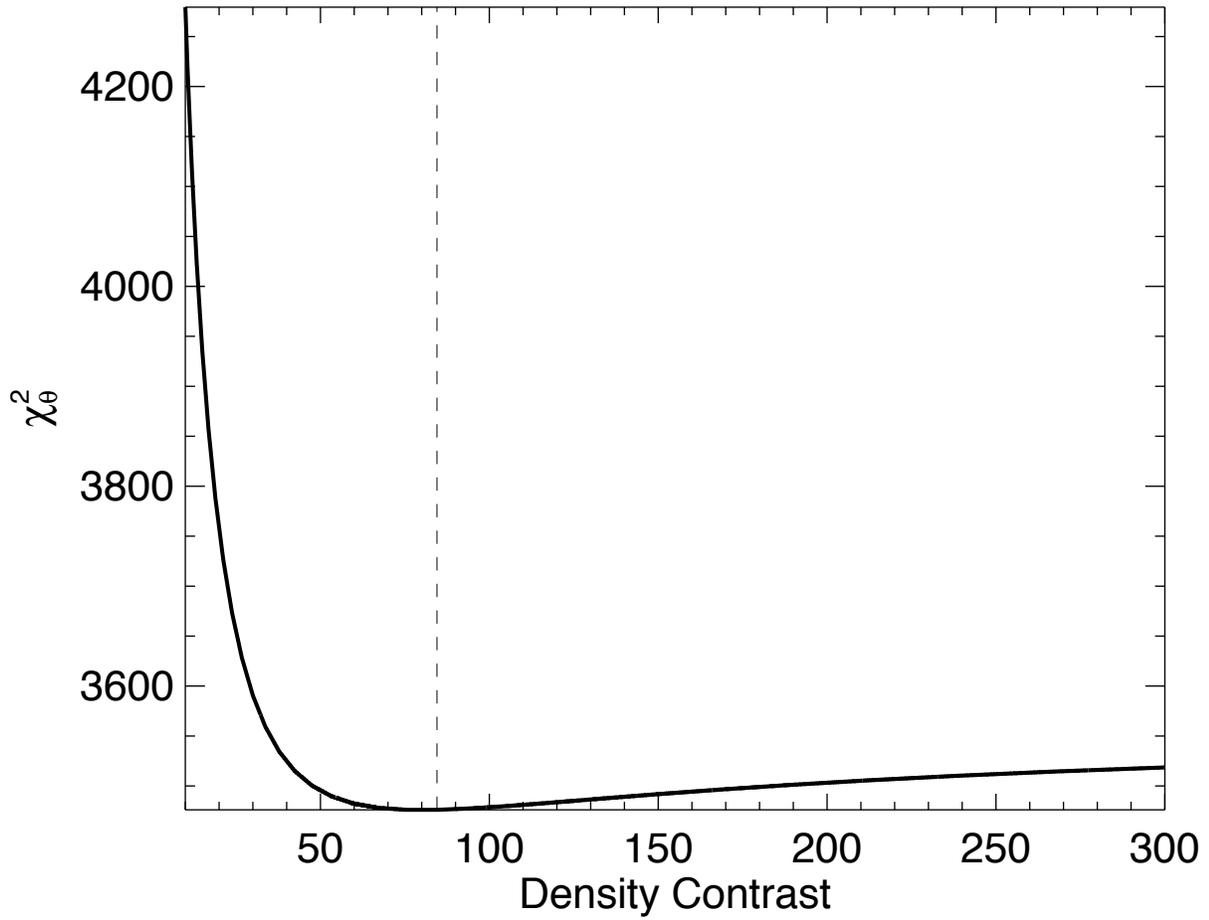}
\end{center}
 \caption{$\chi^2$ distribution of the polarization angle ($\chi_{\theta}^{2}$) against density contrast ($\rho_{\rm c} / \rho_{\rm bkg}$) for the 3D flux freezing model with fixed inclination angle ($\gamma_{\rm mag} = 35^{\circ}$).}
   \label{fig1}
\end{figure}

\clearpage 

\begin{figure}[t]
\begin{center}
 \includegraphics[width=6.5 in]{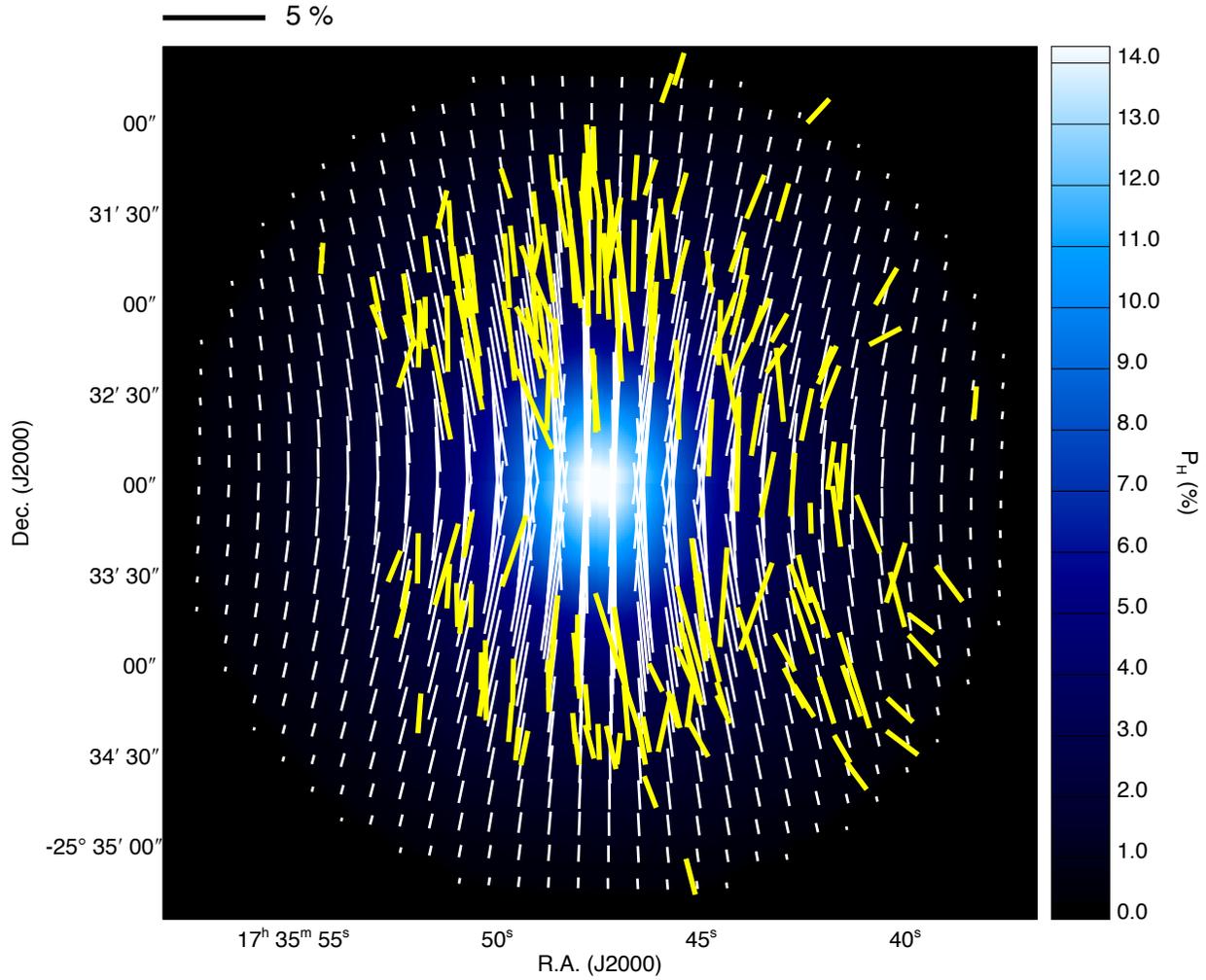}
\end{center}
 \caption{Best-fit 3D flux freezing model ($\gamma_{\rm mag} = 35^{\circ}$ and $\rho_{\rm c} / \rho_{\rm bkg} = 75$, white vectors) with observed polarization vectors (yellow vectors). The background color image shows the polarization degree distribution of the best-fit model. The scale of 5\% polarization degree is shown at the top.}
   \label{fig1}
\end{figure}

\clearpage 

\begin{figure}[t]
\begin{center}
 \includegraphics[width=6.5 in]{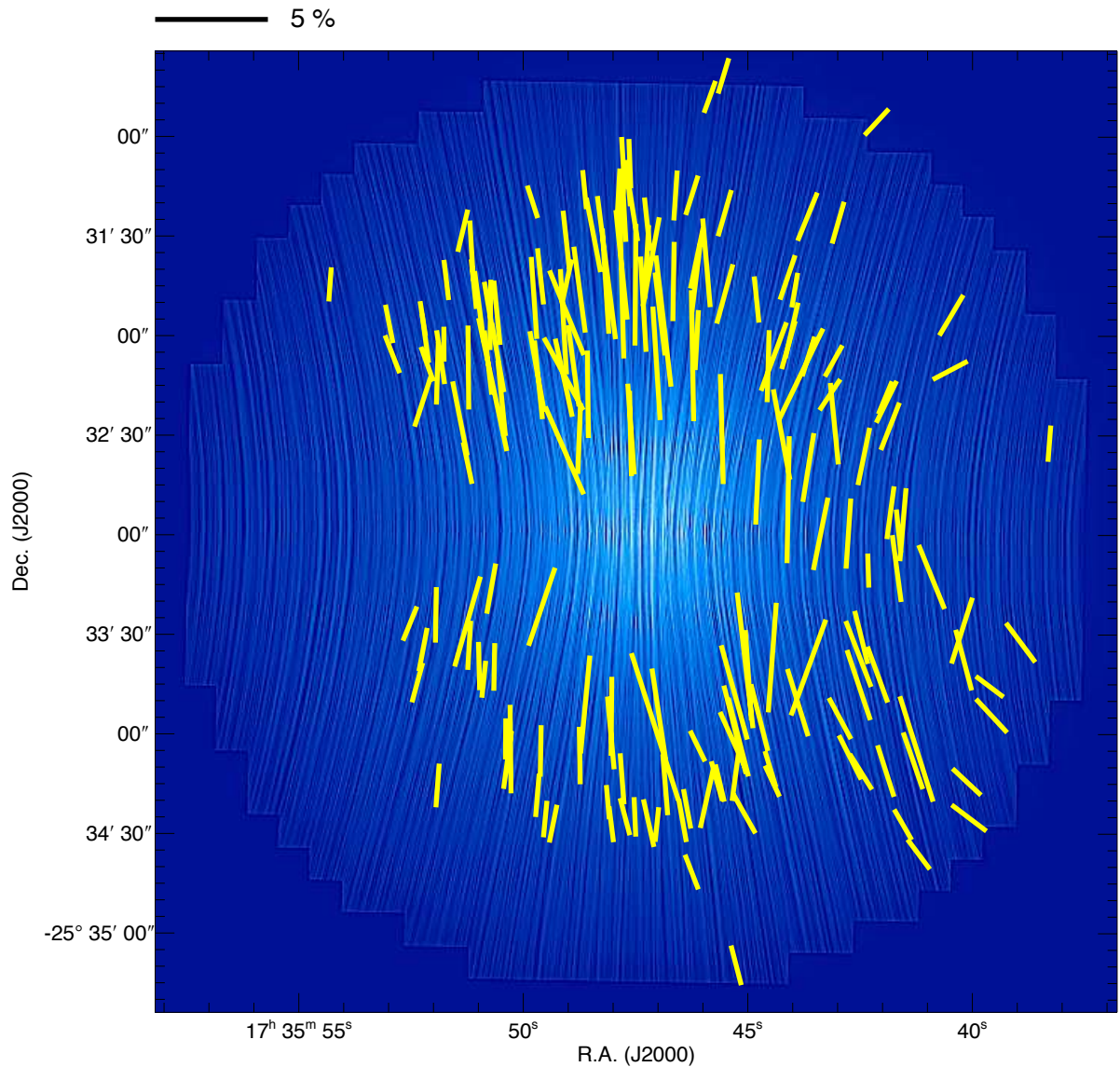}
\end{center}
 \caption{Same as Figure 10, but the background image was made using the line integral convolution technique (LIC: Cabral \& Leedom 1993). The direction of the LIC \lq \lq texture'' is parallel to the direction of the magnetic field, and the background image is based on the polarization degree of the model core.}
   \label{fig1}
\end{figure}

\clearpage 

\begin{figure}[t]
\begin{center}
 \includegraphics[width=6.5 in]{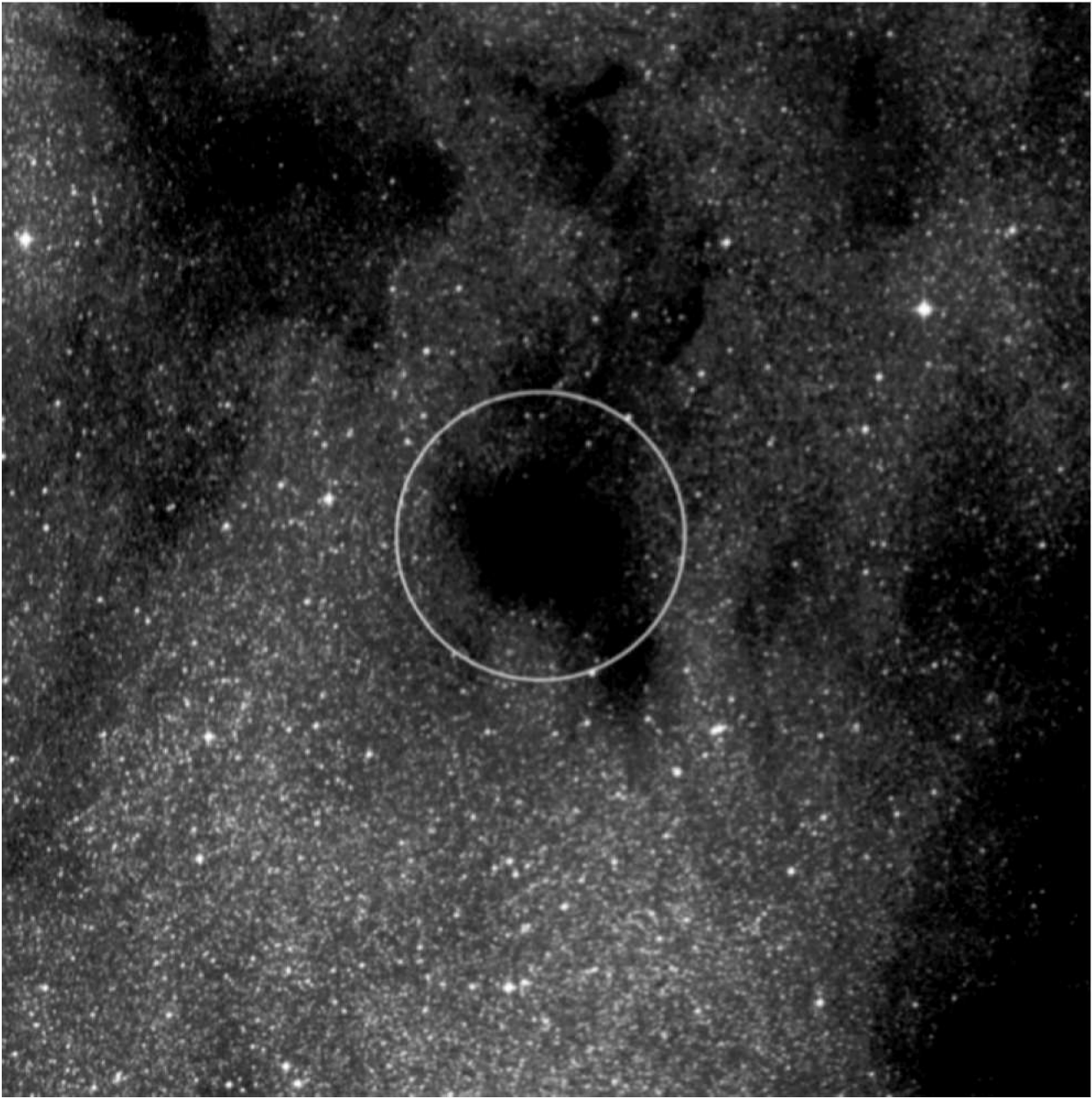}
\end{center}
 \caption{Optical image (Digitized Sky Survey 2, $R$ band) covering a $30'$ extent around FeSt 1-457. The white circle shows the initial radius ($R_0 =1.64 R = 236'' = 0.15$ pc). The optical boundary of the obscuration around the center roughly corresponds to the current radius ($R$) of the core.}
   \label{fig1}
\end{figure}

\clearpage 

\begin{figure}[t]
\begin{center}
 \includegraphics[width=6.5 in]{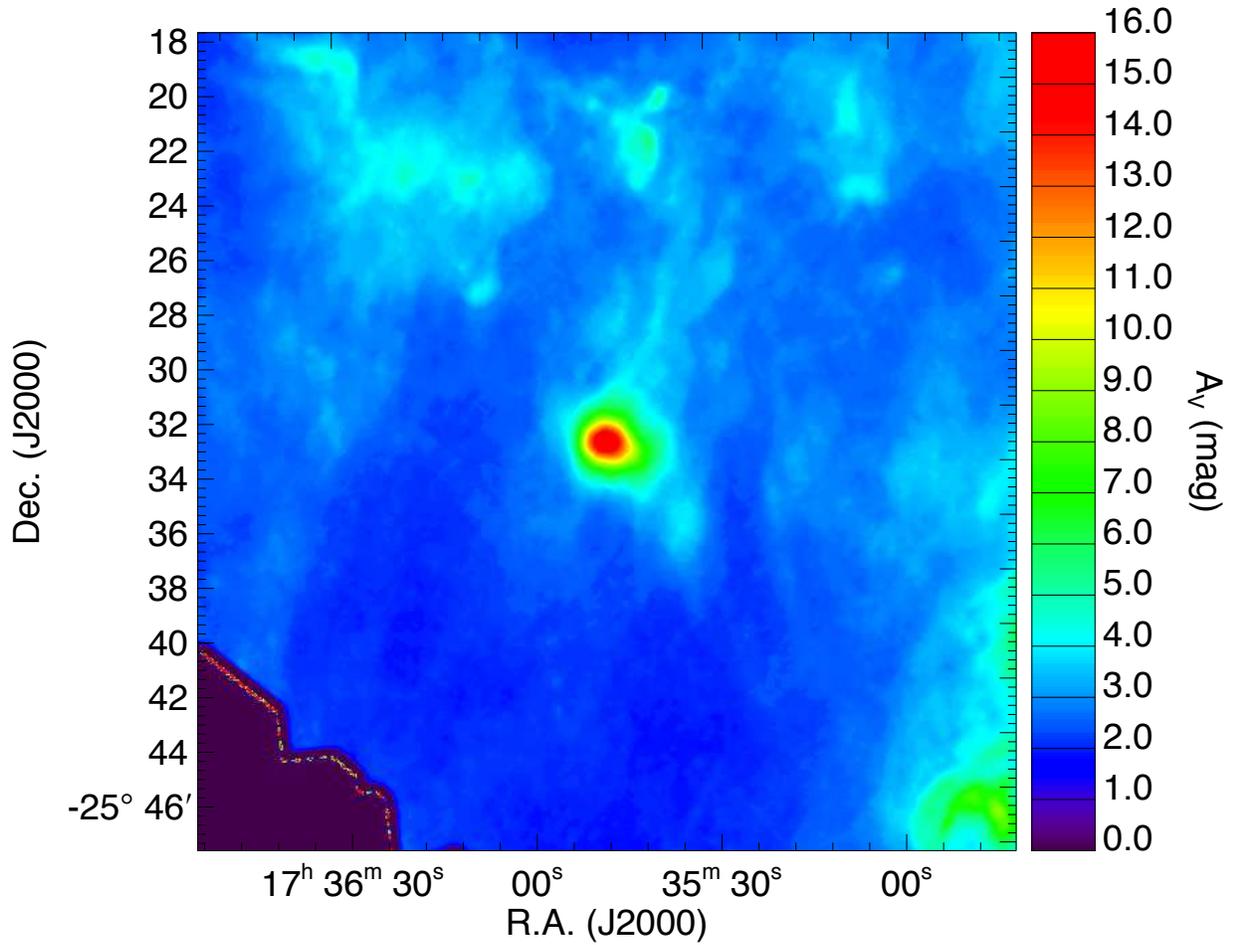}
\end{center}
 \caption{{\it Herschel} column density map (Roy et al. 2019; Andr\'{e} et al. 2010) covering a $30'$ extent around FeSt 1-457. The column density was converted to $A_V$ using $N_{\rm H_2}/A_V = 9.4 \times 10^{20}$ cm$^{-2}$ mag$^{-1}$ (Bohlin, Savage, \& Drake 1978). The resolution of the image is $18.2''$.}
   \label{fig1}
\end{figure}

\clearpage

\beginsupplement

\clearpage 

\begin{figure}[t]
\begin{center}
 \includegraphics[width=6.5 in]{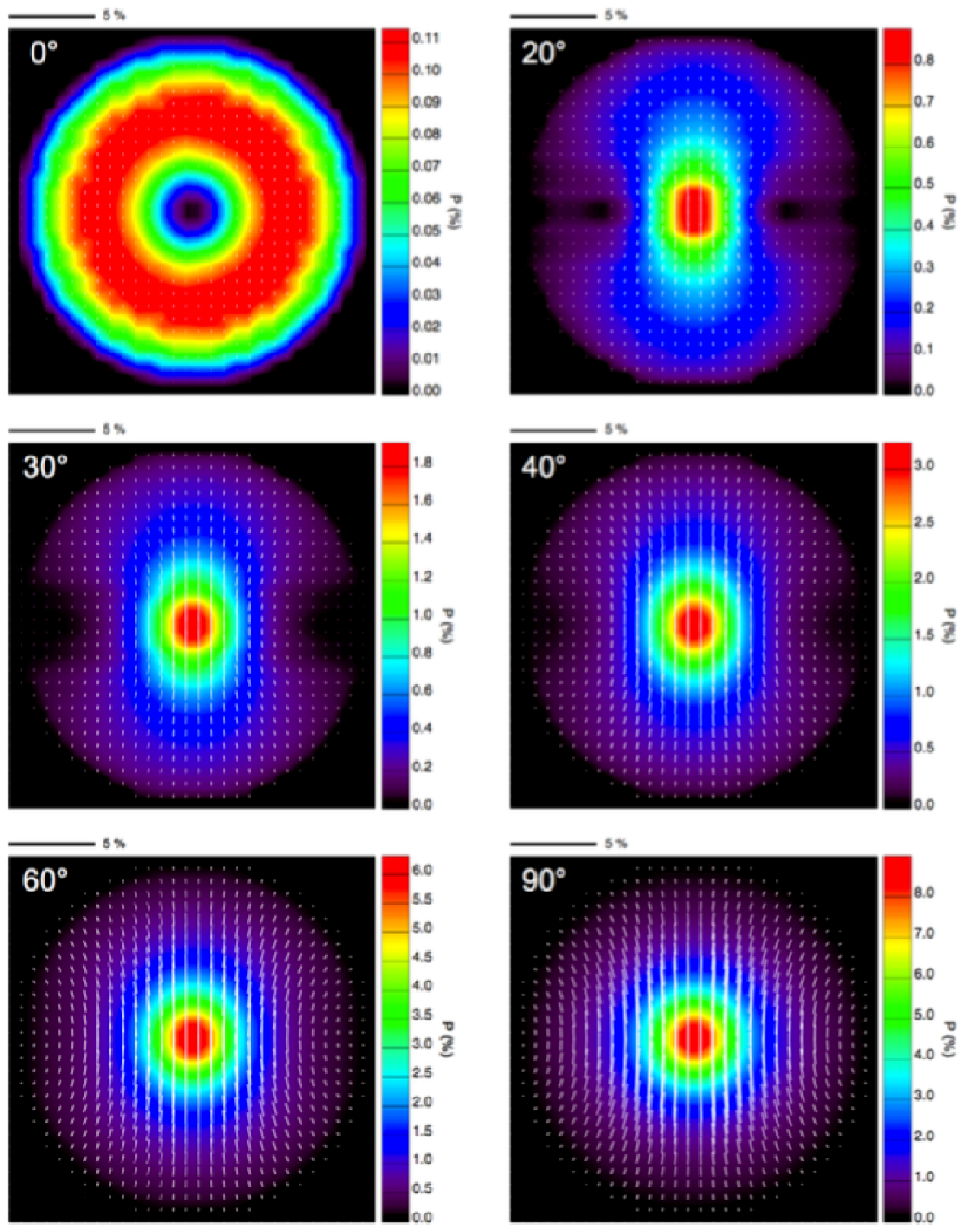}
\end{center}
 \caption{Polarization vector maps of the 3D parabolic model (white vectors). The background color and color bar show the polarization degree. The applied viewing angle ($\gamma_{\rm view} = 90^{\circ}-\gamma_{\rm mag}$) is labeled in the upper left corner of each panel. The magnetic curvature parameter $C$ is set to $2.0 \times 10^{-4}$ arcsec$^{-2}$ for all the panels.}
   \label{fig1}
\end{figure}

\clearpage 

\begin{figure}[t]
\begin{center}
 \includegraphics[width=6.5 in]{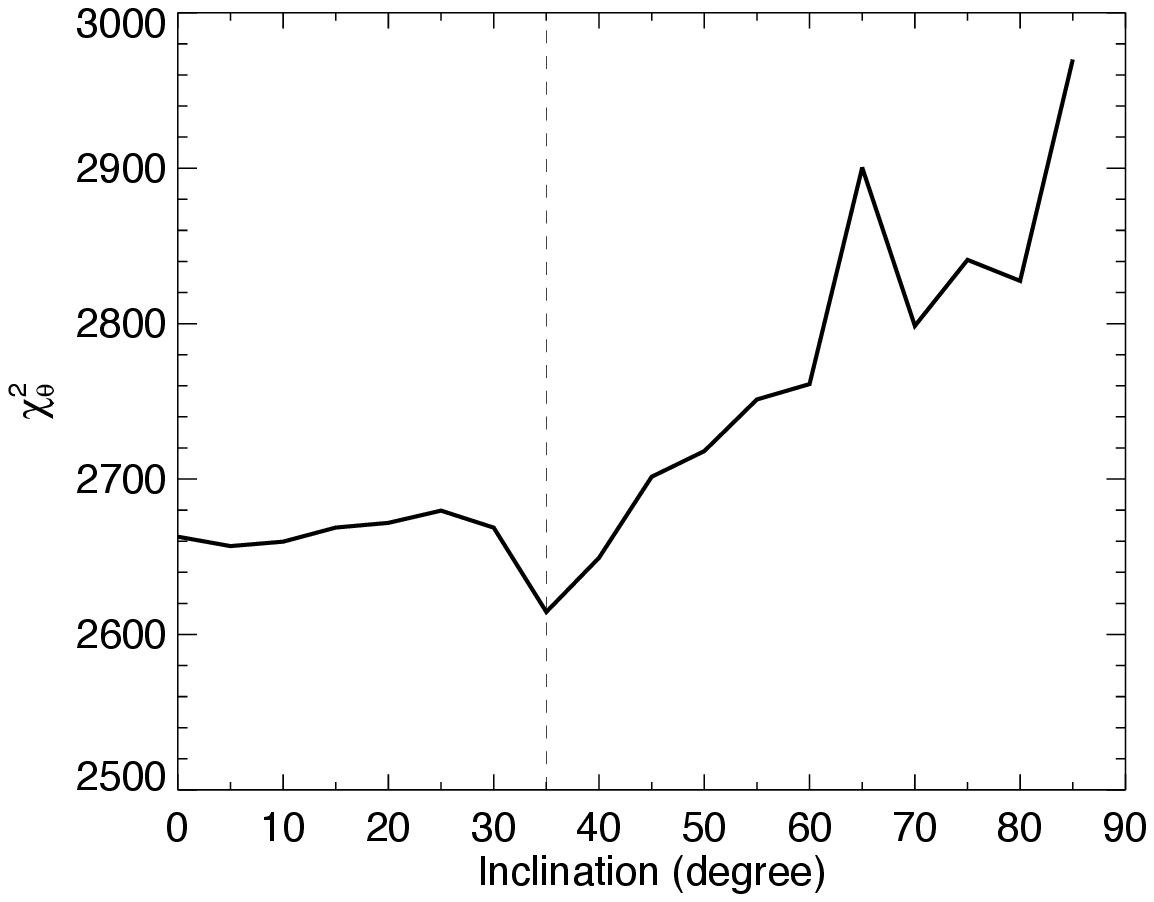}
\end{center}
 \caption{$\chi^2$ distribution of the polarization angle ($\chi^2_{\theta}$). The best magnetic curvature parameter ($C$) was determined for each inclination angle ($\gamma_{\rm mag}$). $\gamma_{\rm mag} = 0^{\circ}$ and $90^{\circ}$ correspond to the edge-on and pole-on geometries with respect to the magnetic axis.}
   \label{fig1}
\end{figure}

\clearpage 

\begin{figure}[t]
\begin{center}
 \includegraphics[width=6.5 in]{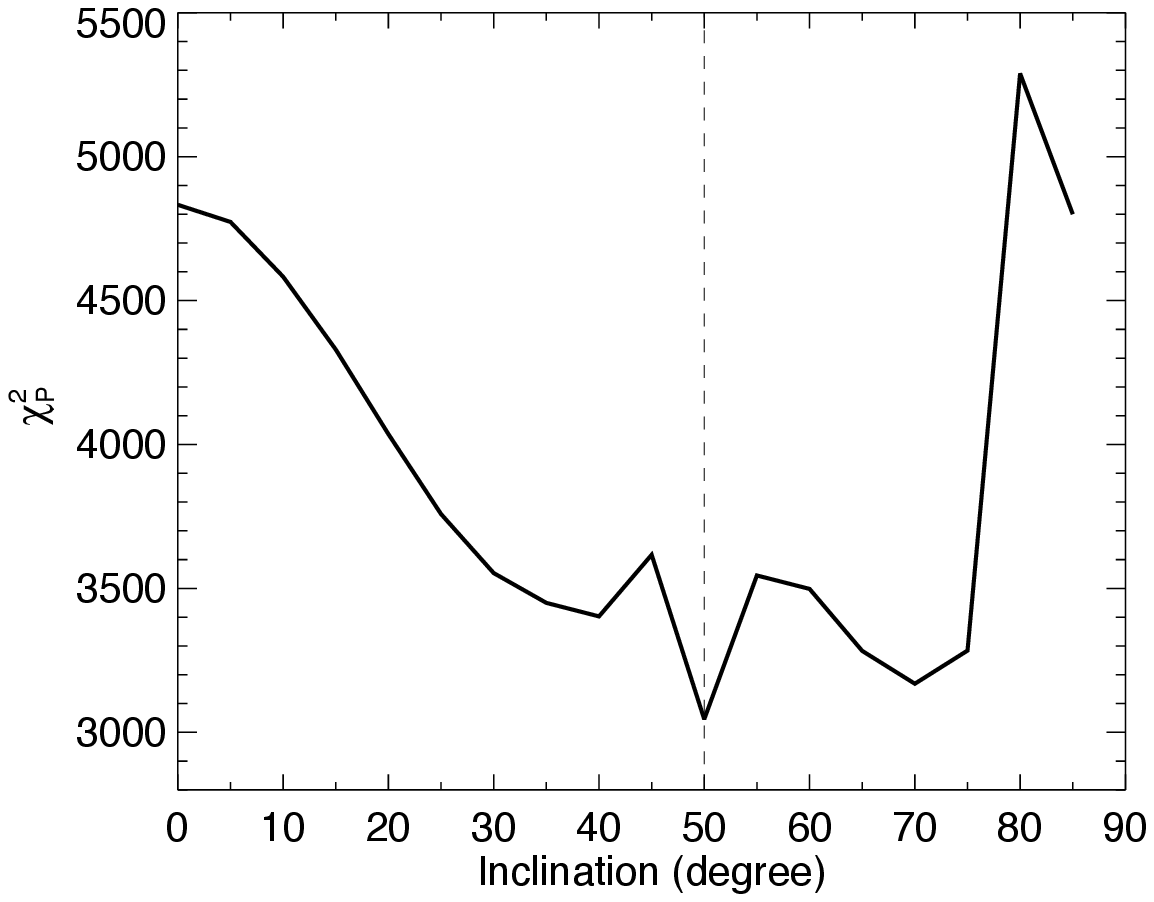}
\end{center}
 \caption{$\chi^2$ distribution of the polarization degree ($\chi^2_P$). The calculations of $\chi^2$ in polarization degree were performed after determining the best magnetic curvature parameter ($C$) that minimized $\chi^2$ in the polarization angle. This calculation was carried out for each inclination angle ($\gamma_{\rm mag}$). $\gamma_{\rm mag} = 0^{\circ}$ and $90^{\circ}$ correspond to the edge-on and pole-on geometries with respect to the magnetic axis.}
   \label{fig1}
\end{figure}

\clearpage 

\begin{figure}[t]
\begin{center}
 \includegraphics[width=6.5 in]{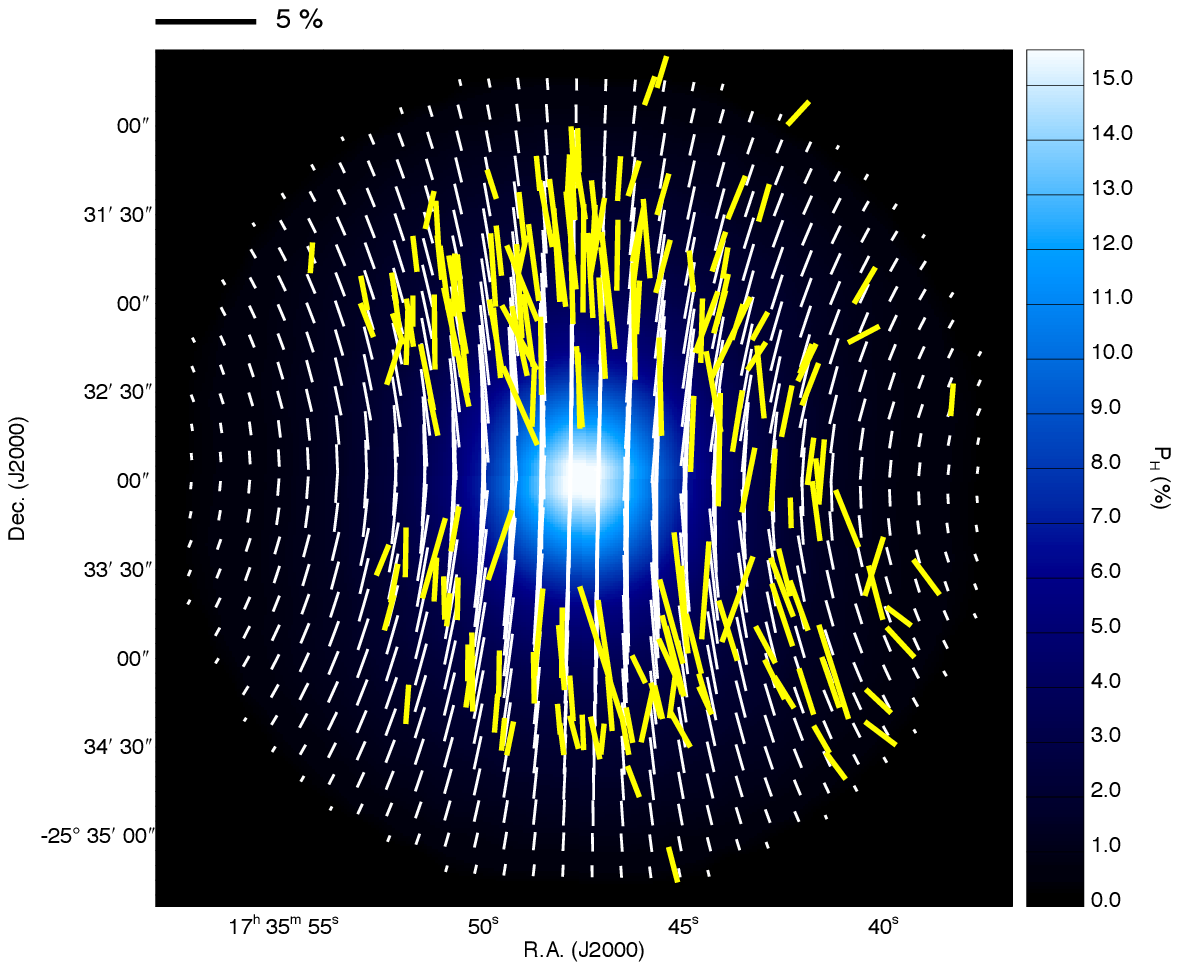}
\end{center}
 \caption{Best-fit 3D parabolic model ($\gamma_{\rm mag} = 35^{\circ}$ and $C = 2.0 \times 10^{-4}$ arcsec$^{-2}$, white vectors) with the observed polarization vectors (yellow vectors). The background color image shows the polarization degree distribution of the best-fit model. The scale of 5\% polarization degree is shown at the top.}
   \label{fig1}
\end{figure}

\clearpage 

\begin{figure}[t]
\begin{center}
 \includegraphics[width=6.5 in]{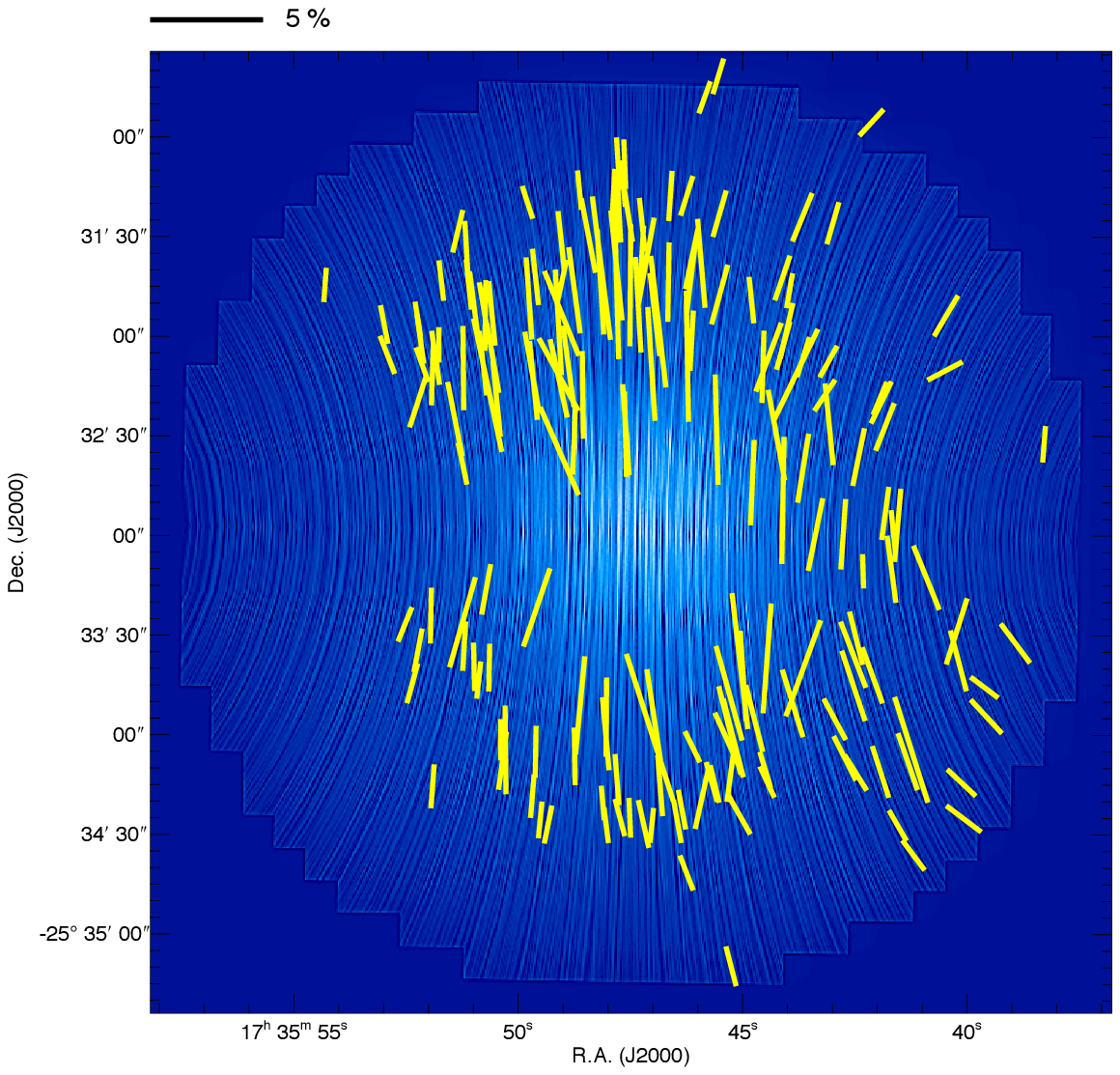}
\end{center}
 \caption{Same as Figure S4, but the background image was made using the line integral convolution technique (LIC: Cabral \& Leedom 1993). The direction of the LIC \lq \lq texture'' is parallel to the magnetic field direction, and the background image is based on the polarization degree of the model core.}
   \label{fig1}
\end{figure}

\clearpage 

\begin{figure}[t]
\begin{center}
 \includegraphics[width=2.5 in]{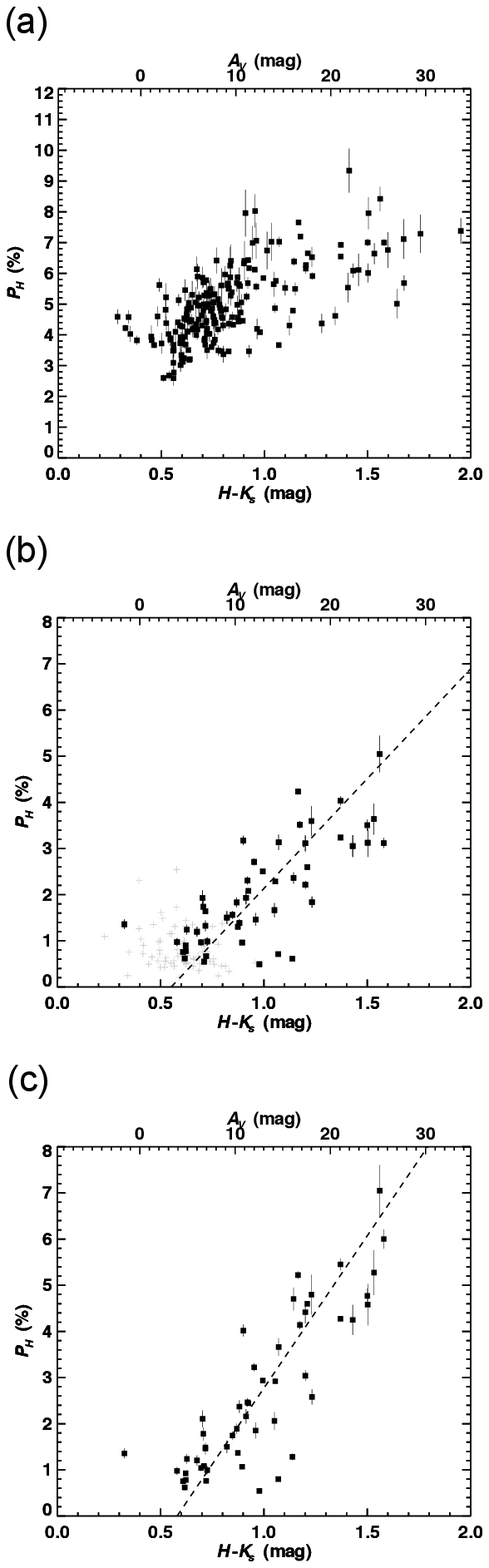}
\end{center}
 \caption{Relationship between polarization degree $P_H$ and $H-K_s$ color toward background stars. Stars with $R \le 144''$ and $P_H / \delta P_H \ge 10$ are plotted. (a) $P$--$A$ relationship with no correction (observed data). (b) $P$--$A$ relationship after correcting for ambient polarization components. The gray plus symbols show the relationship for the stars located in the off-core region ($R > 144''$ and $P_H / \delta P_H \ge 10$). (c) $P$--$A$ relationship after correcting for ambient polarization components, the depolarization effect, and the magnetic inclination angle.}
   \label{fig1}
\end{figure}

\clearpage 

\begin{figure}[t]
\begin{center}
 \includegraphics[width=6.5 in]{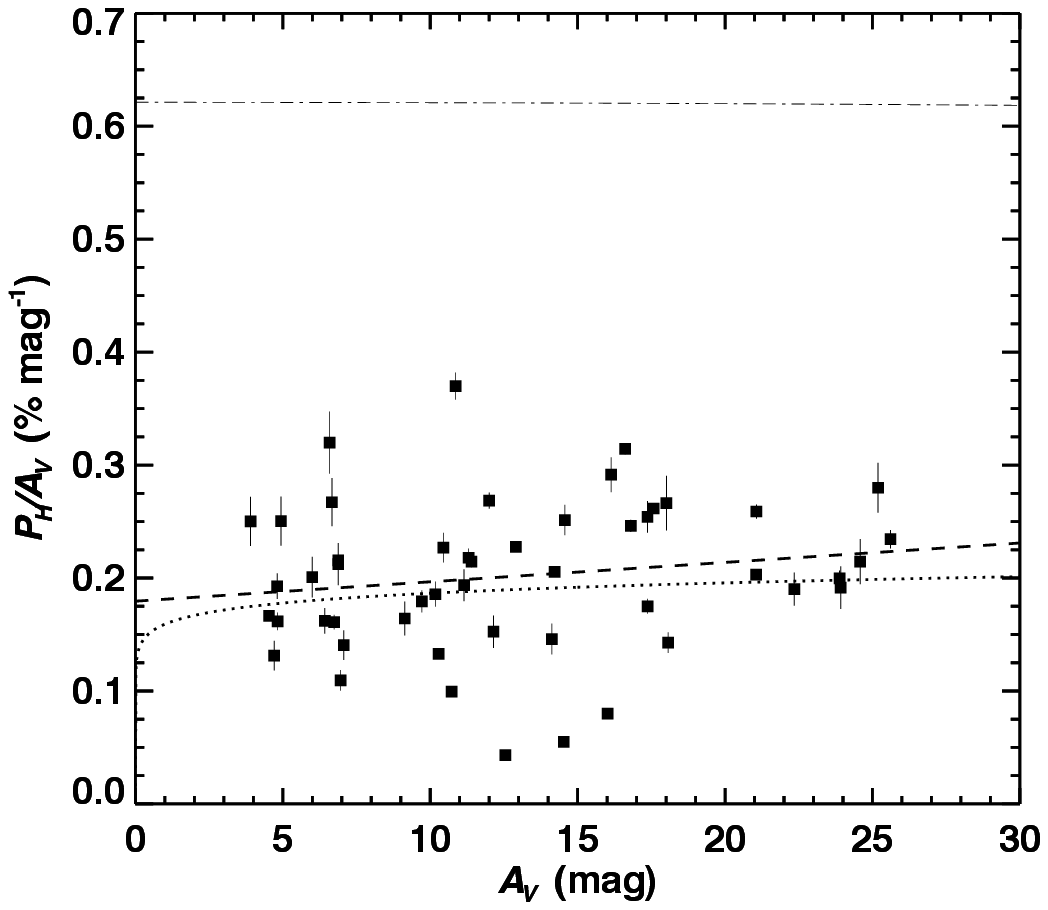}
\end{center}
 \caption{Relationships between $P_H / A_V$ and $A_V$ toward background stars. Stars with $R \le 144''$ and $P_H / \delta P_H \ge 10$ are plotted. The relationship was corrected for ambient polarization components, the depolarization effect, and the magnetic inclination angle. The dashed line denotes the linear least-squares fit to all the data points. The dotted line shows the power-law fitting result. The dotted-dashed line shows the observational upper limit reported by Jones (1989).}
   \label{fig1}
\end{figure}

\clearpage 

\begin{figure}[t]
\begin{center}
 \includegraphics[width=6.5 in]{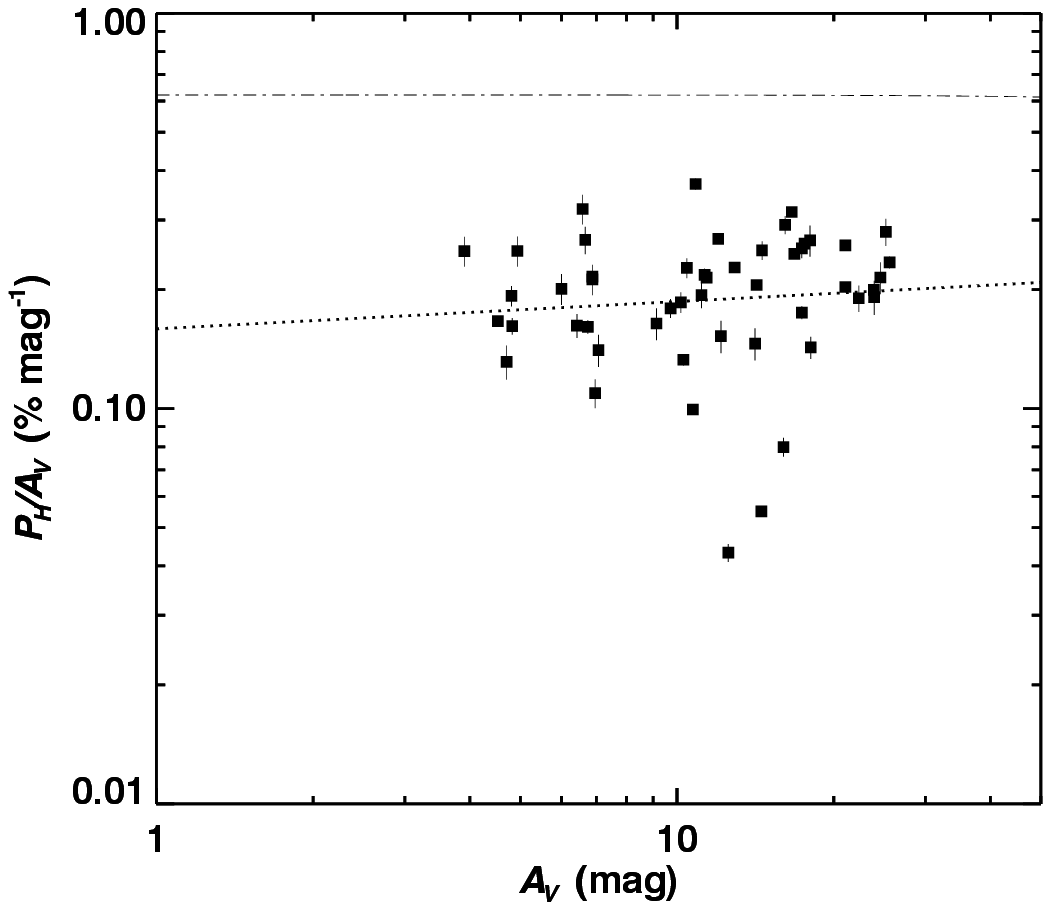}
\end{center}
 \caption{Same as Figure 7, but both axes are shown with logarithmic scales.}
   \label{fig1}
\end{figure}

\clearpage 

\begin{figure}[t]
\begin{center}
 \includegraphics[width=6.5 in]{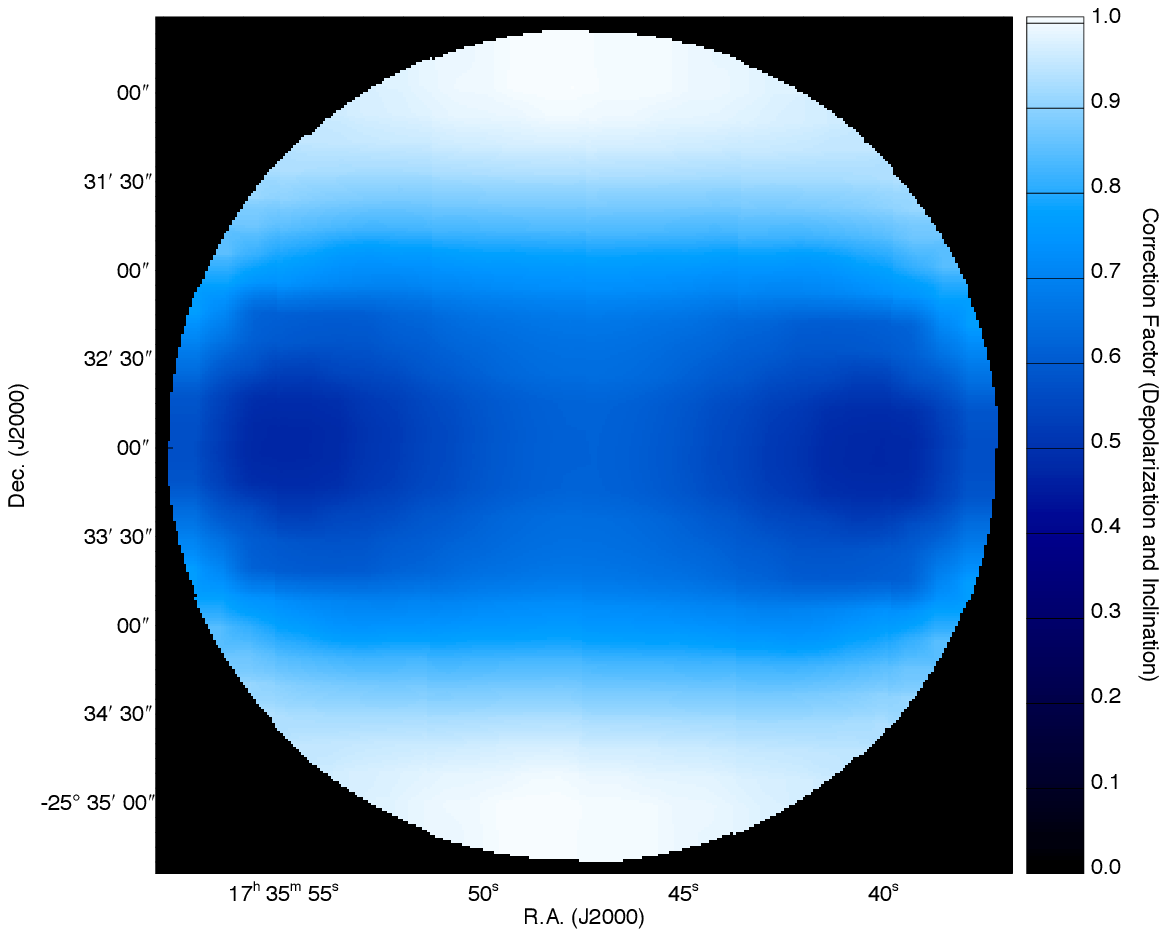}
\end{center}
 \caption{Distribution of the depolarization and inclination correction factor. The field of view is the same as the diameter of the core $288''$.}
   \label{fig1}
\end{figure}

\clearpage 

\begin{figure}[t]
\begin{center}
 \includegraphics[width=6.5 in]{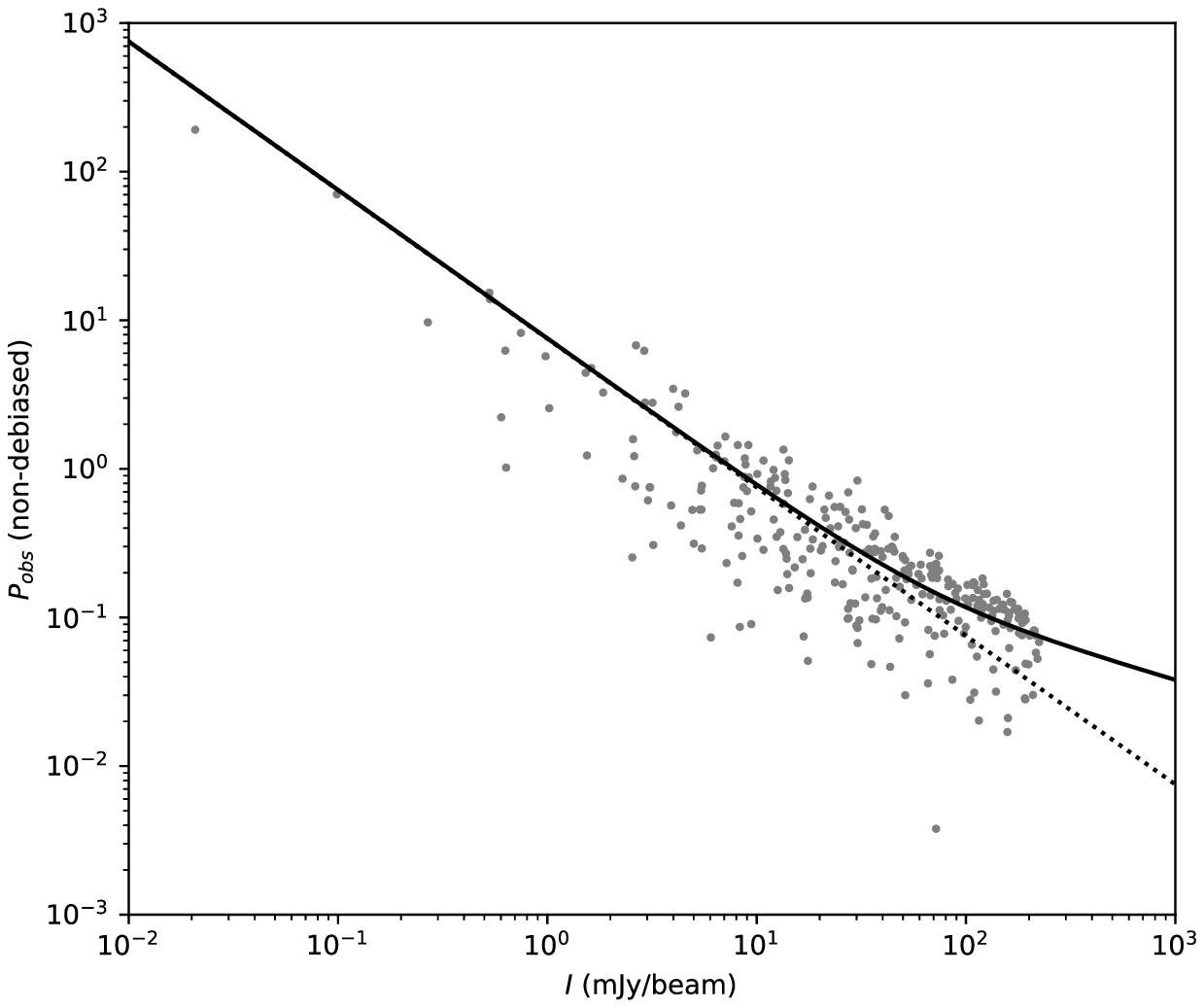}
\end{center}
 \caption{Relationship between the polarization fraction $P_{\rm submm}$ and intensity $I_{\rm submm}$ at submillimeter wavelengths. The solid line shows the best-fitting Ricean-mean model. The dotted line shows the relationship of the low-SNR limit.}
   \label{fig1}
\end{figure}

\clearpage

\fontsize{10pt}{10pt}\selectfont

  \begin{center}
  \begin{longtable}{cccccc} 
  \caption{List of Physical Parameters of FeSt 1-457}
  \label{longtablesample}\\
  \hline \hline
 Symbol            &  Definition            &  Values                        & Units        &  Notes                  & Ref. \\ \hline 
 {\bf Fundamental Param.} &     \           &   \                            &     \        &    \                                  & \  \\ 
 $\alpha$          &  R.A. (J2000)          &  $17^{\rm h}35^{\rm m}47.\hspace{-3pt}^{\rm s}5$ & hms     &  a                       & 1  \\
 $\delta$          &  Decl. (J2000)         &  $-25^{\circ}32'59.\hspace{-3pt}''0$             & dms     &  a                       & 1  \\
 $d$               &  Distance              &  $130^{+24}_{-58}$             & pc           &  b                                    & 2  \\ 
 $\theta_R$        &  Angular radius        &  $144'' \pm 11''$              & arcsec       &  ...                                  & 1  \\ 
 $R$               &  Radius                &  $18500 \pm 1460$              & au           &  $0.093 \pm 0.007$ pc                 & 1  \\ 
 $M$               &  Mass                  &  $3.55 \pm 0.75$               & M$_{\odot}$  &  ...                                  & 1  \\ 
 $\rho_{\rm c}$    &  Density at center     &  $3.50 \pm 0.99 \times 10^5$ & cm$^{-3}$    &  $1.36 \times 10^{-18}$ g cm$^{-3}$     & 1  \\ 
 $\rho_{\rm avg}$  &  Mean density          &  $2.04 \pm 0.65 \times 10^4$ & cm$^{-3}$    &  $7.93 \times 10^{-20}$ g cm$^{-3}$     & 1  \\ 
 $P_{\rm ext}$     &  External pressure     &  $1.1 \pm 0.3 \times 10^5$   & K cm$^{-3}$  &    c                                    & 1  \\ 
 $\xi_{\rm max}$   &  Nondimensional radius &  $12.6 \pm 2.0$                & ...          &  d                                    & 1  \\ 
 $\rho_{\rm c}/\rho_{\rm s}$ &  Density contrast &  74.5                     & ...          &  e                                    & 1  \\ 
 $A_{V,{\rm center}}$ & $A_V$ toward center &  41                            & mag          &  f                                    & 1  \\ 
 $T_{\rm kin}$     &  Kinematic temperature &  $9.5$                         & K            &  g                                    & 3  \\ 
 $V_{\rm LSR}$     &  Line center velocity  &  $5.820 \pm 0.003$             & km s$^{-1}$  &  h                                    & 1  \\ 
 $\Delta V$        &  FWHM line width       &  $0.182 \pm 0.006$             & km s$^{-1}$  &  h                                    & 1  \\ 
 $\sigma_{\rm turb}$ &  Turbulent velocity dispersion &  $0.0573 \pm 0.006$  & km s$^{-1}$  &  h                                    & 1  \\ 
 $\delta V_{\rm c}$  &  Centroid velocity dispersion  &  $0.023$             & km s$^{-1}$  &  h                                    & 1  \\ 
 $M_{\rm BE}$      &  Bonnor--Ebert mass    &  $1.19 \pm 0.32$               & M$_{\odot}$  &  ...                                  & 1  \\ 
 $E_{\rm rotation}/E_{\rm gravity}$ & Energy ratio  &  $\sim 0.01$           & ...          &  i                                    & 4  \\ 
 ...               & Gas infalling motion?  &  No                            & ...          &  j                                    & 4  \\ 
 ...               & Association of YSOs?   &  None                          & ...          &  k                                    & 5,6\\ 
 $\theta_{\rm rot}$  & Core's rotation axis   &  $140^{\circ}-160^{\circ}$   & degree       &  l                                    & 4  \\ 
 $\theta_{\rm elon}$ & Core's elongation axis &  $\sim 90^{\circ}$           & degree       &  ...                                  & 1  \\ 
 \hline
 {\bf Magnetic Param.} &    \               &     \                              &   \            &          \                      & \    \\ 
 $\theta_{\rm mag}$ & Magnetic axis (pos) &  $179^{\circ} \pm 11^{\circ}$ & degree          &  m                                    & 7  \\ 
 $\gamma_{\rm mag}$ & Magnetic axis (los) &  $35^{\circ} \pm 15^{\circ}$ & degree           &  n                                    & 8,13\\ 
 $C_{\rm 2D}$  &  Magnetic curvature (2D) &  $5.14 \pm 2.22 \times 10^{-5}$ & arcsec$^{-2}$ &  o                                    & 7  \\ 
 $C_{\rm 3D}$  &  Magnetic curvature (3D) &  $2.01 \times 10^{-4}$ & arcsec$^{-2}$          &  o                                    & 8,13\\ 
 $B_{\rm pos}$ &  B-field strength (pos)  &  $23.8 \pm 12.1$ & $\mu$G                       &  p                                    & 7  \\ 
 $B_{\rm tot}$ &  B-field strength (total)&  $28.9 \pm 14.8$ & $\mu$G                       &  q                                    & 8,13\\ 
 $B_{\rm tot,edge}$   & $B_{\rm tot}$ at core edge &  $14.6$ & $\mu$G &  ...                                                        & 10,13\\ 
 $B_{\rm tot,center}$ & $B_{\rm tot}$ at core center & $113.5$ & $\mu$G &  ...                                                      & 10,13\\ 
 $\lambda_{\rm pos}$ & Mass-to-flux ratio (pos) &  $2.00$               & ...               &  r                                    & 7   \\ 
 $\lambda_{\rm tot}$ & Mass-to-flux ratio (total) &  $1.64$             & ...               &  r                                    & 8,13\\ 
 $\lambda_{\rm tot, edge}$   & $\lambda_{\rm tot}$ at core edge & $\approx 1$   & ...       &  r                                    & 10,13\\ 
 $\lambda_{\rm tot, center}$ & $\lambda_{\rm tot}$ at core center & $\approx 2$ & ...       &  r                                    & 10,13\\ 
 $M_{\rm mag}$ &  Magnetic critical mass  & $2.16 \pm 0.65$ & M$_{\odot}$ &  ...                                                    & 8,13\\ 
 $M_{\rm cr}$  &  Critical mass ($M_{\rm mag}+M_{\rm BE}$) & $3.35 \pm 0.75$ & M$_{\odot}$ &  ...                                   & 8,13\\ 
 $\kappa$      &  B-field scaling ($|B|\propto \rho^{\kappa}$) & $0.78 \pm 0.10$ & ...          &  ...                              & 10\\ 
 $\beta$            &  Energy ratio ($3C_{s}^{2}/V_{\rm Alfv\acute{e}n}^2$)  & $1.27$ & ...                  &  s                   & 8,13\\ 
 $\beta_{\rm turb}$ &  Energy ratio ($3\sigma_{\rm turb}^{2}/V_{\rm Alfv\acute{e}n}^2$) & $0.12$ &  ...      &  s                   & 8,13\\ 
 \hline
 {\bf NIR Polarimetry} &   \             &      \                         &    \         &    \                                     & \  \\ 
 $\delta \theta_{\rm int}$ & Polarization angle dispersion& $6.90^{\circ} \pm 2.72^{\circ}$ & degree         &  t                   & 7  \\  
 $(P_H / E_{H-K_s})_{\rm obs}$ & Polarization efficiency  & $2.43 \pm 0.05$ & $\%$ mag$^{-1}$                &  u                   & 7,9\\ 
 $(P_H / E_{H-K_s})_{\rm bkg}$ & Polarization efficiency  & $4.76 \pm 0.33$ & $\%$ mag$^{-1}$                &  v                   & 7,9\\ 
 $(P_H / E_{H-K_s})_{\rm all}$ & Polarization efficiency  & $6.60 \pm 0.41$ & $\%$ mag$^{-1}$                &  w                   & 9,13\\ 
 $(P_H / A_V) / A_V$           & Polarization efficiency  & $0.002 \pm 0.002$ & $\%$ mag$^{-1}$              &  x                   & 11,13\\ 
 $\alpha_{\rm H}$ & $P_H / A_V \propto A_{V}^{-\alpha_{\rm H}}$ & $-0.07 \pm 0.11$ & ...                     &  x                   & 11,13\\ 
 \hline
 {\bf Submm Polarimetry} &   \             &    \                           &   \          &    \                                   & \  \\ 
 $\theta_{\rm mag,submm}$ & Magnetic axis (pos,submm) & $132.1^{\circ} \pm 22.0^{\circ}$ & degree &  ...                            & 11,12 \\ 
 $\alpha_{\rm Rice,submm}$ & $P_{\rm submm} \propto I^{-\alpha_{\rm Rice,submm}}$ & $0.41 \pm 0.10$    & ...          &  y          & 11,12,13\\ 
 \hline
 {\bf Core Formation} &   \             &    \                           &   \          &    \                                      & \  \\ 
 $\rho_{\rm 0}$          &  Initial density            &  4670               & cm$^{-3}$       &  $1.82 \times 10^{-20}$ g cm$^{-3}$& 13\\ 
 $\theta_{R_0}$          &  Initial angular radius     &  $236''$            & arcsec          &  ...                               & 13\\ 
 $R_{\rm 0}$             &  Initial radius             &  $3.0 \times 10^4$  & au              &  $0.15$ pc                         & 13\\ 
 $B_{\rm 0}$             &  Initial B-field Strength   &  $10.8$--$14.6$     & $\mu$G          &  ...                               & 13\\ 
 $M_{J,{\rm ini}}$       &  Jeans mass (initial)       &  $4.21$             & M$_{\odot}$     &  ...                               & 13\\ 
 $\lambda_{J,{\rm ini}}$ &  Jeans length (initial)     &  $6.4 \times 10^4$  & au              &  $0.32$ pc                         & 13\\ 
 $t_{\rm ff,ini}$        &  Free-fall time             &  $4.9 \times 10^5$  & yr              &  z                                 & 13\\ 
 $t_{\rm sc,ini}$        &  Sound crossing time        &  $1.5 \times 10^6$  & yr              &  z                                 & 13\\ 
 \hline
  \end{longtable} 
  \end{center}
\ \\
$^{\rm a}$ The centroid center of the core measured on the $A_V$ map. $^{\rm b}$ Alves \& Franco (2007) estimated the distance to the Pipe Nebula to be $145 \pm 16$ pc based on optical polarimetry. Dzib et al. (2018) estimated the distance to the Barnard 59 (B59) cloud in the Pipe Nebula to be $163 \pm 5$ pc based on the GAIA data. 
$^{\rm c}$ The $P_{\rm ext}$ value was taken from Table 5 of Kandori et al. (2005). The value was determined based on the assumption that the Bonnor--Ebert equilibrium is maintained. However, $P_{\rm ext} = 1.1 \times 10^{5}$ K cm$^{-3}$ is larger than $(T_{\rm kin}+T_{\rm turb}) \times \rho_{\rm c} / 75 \approx 4.9 \times 10^{4}$ K cm$^{-3}$, where $T_{\rm turb}$ is the temperature equivalent to the turbulent velocity dispersion. The latter external pressure value is based on a distance of $130$ pc and a density contrast of $75$ calculated from $\xi_{\rm max} = 12.6$. We chose the former value in the present study. If we use the latter value, the Bonnor--Ebert mass of $M_{\rm BE} = 1.15 \times ((T_{\rm kin}+T_{\rm turb})/10)^2/(P_{\rm ext}/10^5)^{1/2}$ (McKee 1999) is $1.79$ M$_{\odot}$, and $M_{\rm cr}$ increases to $3.95$ M$_{\odot}$. Comparing the observed core mass with $M_{\rm cr}$, the core is still located in a nearly critical state, and the conclusions of this paper do not change. 
$^{\rm d}$ This parameter serves as a stability criterion of the Bonnor--Ebert sphere (Ebert 1955; Bonnor 1956). $^{\rm e}$ The density contrast is the value of the central density $\rho_{\rm c}$ divided by the surface density $\rho_{\rm s}$. $^{\rm f}$ This value was measured on the $A_V$ map with a resolution of $33''$ (Kandori et al. 2005). $^{\rm g}$ Measured using the rotation temperature of the NH$_3$ molecule (Rathborne et al. 2008). $^{\rm h}$ Measured using the N$_2$H$^+$ ($J=1-0$) molecular line (Kandori et al. 2005). $^{\rm i}$ The ratio of rotational energy and gravitational energy. $^{\rm j}$ Aguti et al. (2007) suggests the existence of oscillation in the outer gas layer of FeSt 1-457. $^{\rm k}$ Forbrich et al. (2009,2010) searched young stars in the Pipe Nebula region in the mid-infrared and X-ray wavelengths, and no young sources were found toward the FeSt 1-457 core. $^{\rm l}$ Measured using the N$_2$H$^+$ ($J=1-0$) molecular line (Aguti et al. 2007). $^{\rm m}$ The plane of sky inclination angle of the core's magnetic axis was measured after subtracting the ambient polarization vector component (Paper I)., $^{\rm n}$ Though the line of sight inclination angle of the core's magnetic axis (measured from the plane of sky) was previously estimated to be $45^{\circ} \pm 10^{\circ}$, the value was updated in this paper to $35^{\circ} \pm 15^{\circ}$. $^{\rm o}$ The magnetic curvature term $C$ was used in the simple parabolic magnetic field model, $y=gCx^2$, and its 3D version (Paper I,II). $^{\rm p}$ The plane of sky magnetic field strength estimated using the Davis--Chandrasekhar--Fermi method (Davis 1951; Chandrasekhar \& Fermi 1953). $^{\rm q}$ The total magnetic field strength obtained by dividing $B_{\rm pos}$ by $\cos \gamma_{\rm mag}$. $^{\rm r}$ The mass-to-flux ratio is defined as the observed ratio divided by the theoretical critical value: $\lambda = (M/\Phi)_{\rm obs} / (M/\Phi)_{\rm critical}$. We used $1/2\pi G^{1/2}$ (Nakano \& Nakamura 1978) for the critical value. $^{\rm s}$ The speed of sound at 9.5 K ($C_s$), the turbulent velocity dispersion ($\sigma_{\rm turb}$), and the Alfv\'{e}n velocity were used to estimate the ratios between the thermal, turbulent, and magnetic energies. $^{\rm t}$ $\delta \theta_{\rm int}$ was derived from $\delta \theta_{\rm int} = (\delta \theta_{\rm res}^2 - \delta \theta_{\rm err}^2)^{1/2}$, where $\delta \theta_{\rm err}$ is the observational error in the polarization measurements and $\delta \theta_{\rm res}$ is the standard deviation of the residual angle $\theta_{\rm res} = \theta_{\rm obs} - \theta_{\rm fit}$. The residual angle is obtained by subtracting $\theta_{\rm fit}$, the fitted angle using the parabolic function $y=gCx^2$, from the observed polarization angle $\theta_{\rm obs}$. 
$^{\rm u}$ The polarization efficiency was measured using the observed data with no correction. $^{\rm v}$ The polarization efficiency was measured after subtracting the ambient (off-core) polarization component from the polarizations of the core's background stars. $^{\rm w}$ The polarization efficiency was estimated after three corrections: 1) subtraction of the ambient (off-core) polarization component, 2) correction of depolarization due to the distorted, inclined polarization structure, 3) correction of the effect of line of sight inclination of the magnetic axis. $^{\rm x}$ $P_H$ data after the above three corrections was used. $^{\rm y}$ Following Pattle et al. (2018), all the submillimeter polarization data points without debiasing were used for the fitting with the Ricean-mean model in to estimate the power-law index, $\alpha_{\rm Rice, submm}$. $^{\rm z}$ Calculated based on the initial density $\rho_0$ and the initial radius $R_0$. 
\\
References: (1) Kandori et al. (2005), (2) Lombardi et al. (2006), (3) Rathborne et al. (2008), (4) Aguti et al. (2007), (5) Forbrich et al. (2009), (6) Forbrich et al. (2010), (7) Kandori et al. (2017a/Paper I), (8) Kandori et al. (2017b/Paper II), (9) Kandori et al. (2018a/Paper III), (10) Kandori et al. (2018b/Paper IV), (11) Kandori et al. (2018c/Paper V), (12) Alves et al. (2014,2015), (13) This paper.

\end{document}